\documentclass[fleqn,usenatbib]{mnras}

\usepackage{newtxtext,newtxmath}

\usepackage[T1]{fontenc}
\usepackage{ae,aecompl}

\usepackage{graphicx}	
\usepackage{amsmath}	
\usepackage{amssymb}	
\usepackage{subfigure}
\usepackage{xcolor}

\newcommand{\del}{\partial}
\newcommand{\curl}{\nabla\times }
\newcommand{\dv}{\nabla\cdot}

\newcommand{\ephi}{{\boldsymbol{\hat{\phi}}}}
\newcommand{\ez}{{\bf{\hat{z}}}}
\newcommand{\tomega}{\tilde{\omega} }
\newcommand{\tA}{\textrm{A}}

\newcommand{\ti}{\textrm{i}}
\newcommand{\td}{\textrm{d}}

\allowdisplaybreaks

\title[Toroidal fields and trapped waves]{Quasi-periodic oscillations, trapped inertial waves and strong toroidal magnetic fields in relativistic accretion discs}

\author[J. Dewberry et al.]{
Janosz W. Dewberry$^{1}$,\thanks{E-mail: jwd43@cam.ac.uk }
Henrik N. Latter$^{1}$,
Gordon I. Ogilvie$^{1}$\\
$^{1}$DAMTP, University of Cambridge, CMS, Wilberforce Road, Cambridge, CB3 0WA, UK
}

\date{Accepted XXX. Received YYY; in original form ZZZ}

\pubyear{2018}

\begin{document}
\label{firstpage}
\pagerange{\pageref{firstpage}--\pageref{lastpage}}
\maketitle

\begin{abstract}
The excitation of trapped inertial waves (r-modes) by warps and eccentricities in the inner regions of a black hole accretion disc may explain the high-frequency quasi-periodic oscillations (HFQPOs) observed in the emission of Galactic X-ray binaries. However, it has been suggested that strong vertical magnetic fields push the oscillations' trapping region toward the innermost stable circular orbit (ISCO), where conditions could be unfavourable for their excitation. This paper explores the effects of large-scale magnetic fields that exhibit \textit{both} toroidal and vertical components, through local and global linear analyses. We find that a strong toroidal magnetic field can reduce the detrimental effects of a vertical field: in fact, the isolation of the trapping region from the ISCO may be restored by toroidal magnetic fields approaching thermal strengths. The toroidal field couples the r-modes to the disc's magneto-acoustic response and inflates the effective pressure within the oscillations. As a consequence, the restoring force associated with the vertical magnetic field's tension is reduced. Given the analytical and numerical evidence that accretion discs threaded by poloidal magnetic field lines develop a strong toroidal component, our result provides further evidence that the detrimental effects of magnetic fields on trapped inertial modes are not as great as previously thought. 

\end{abstract}




\section{Introduction}
High frequency quasi-periodic oscillations (HFQPOs), observed in the light curves of Galactic black hole binaries (BHBs), are a striking but poorly understood phenomenon. Appearing as coherent peaks in the power density spectrum (PDS) of these sources, HFQPOs have aroused particular interest because their frequencies (of $\sim 50-500$Hz) are (a) comparable to the characteristic orbital and epicyclic frequencies of the inner accretion flow, (b) inversely related to black hole mass (when known), and (c) relatively insensitive to substantial variations in luminosity. This suggests that they are connected to the intrinsic properties of the central black hole, and may provide further means of probing the structure of strongly curved space time \citep{Rem06}.

The first HFQPO found in a BHB was a transient $67$Hz oscillation observed in the PDS of GRS 1915+105 \citep{Morg97}. This persistently active system exhibits frequent HFQPOs, but confirmed observations are far less common in other BHBs: in fact, there have only been reliable detections in 5-10 sources. Significantly, HFQPOs appear only in outbursting states in which the flux and inferred accretion rates are exceptionally high, the so-called `Steep Power Law' (SPL) or `very high' state. The properties of the accretion disc during this phase are not well constrained. In particular, it is unclear if the disc extends to the innermost stable circular orbit (ISCO) or truncates before then in a hot torus \citep{Done07}. To complicate matters further, some BHBs exhibit two HFQPOs with different frequencies. These usually appear at different times, but have been observed simultaneously in GRO J1655-40 \citep{Bel12,Mot14}. The phenomenology of HFQPOs is rich, and in some cases connected to the more prevalent low frequency QPOs (LFQPOs). For reviews the reader might consult \citet{Rem06}, \citet{Done07}, and \citet{Mot16}.

Most of the theories offered as an explanation for HFQPOs are dynamical, with radiative and thermal physics yet to be explored in great detail \citep[but see, for example,][]{DexBl,Cab}. Moreover, many appeal to test particle dynamics. For example, the observation that multiple HFQPOs often appear with frequencies in ratios near 3:2 led to models appealing to resonances at special radii where the orbital and epicyclic frequencies achieve the same commensurability \citep{Kluz01,Abr01}. Separate but related is the `relativistic precession model' (RPM), which associates the orbital, apsidal precession, and nodal (or Lense-Thirring) precession frequencies of particles near a Kerr black hole with an upper HFQPO, a lower HFQPO and a Type-C LFQPO, respectively \citep{Stel98,Stel99,Mot14b,Mot17}. Neither model offers a robust explanation for how these dynamical features might lead to large amplitude modulations in emissivity. 

An accretion disc, however, is not a collection of non-interacting particles, and so the mentioned theories must map particle oscillations onto global fluid dynamical waves. This can be achieved, to some extent, if the accretion flow is regarded as a slender torus of constant angular momentum \citep{Rez03,Blaes06,Horak08,Frag16}. In this case, one must assume that the geometry of the inner accretion flow is indeed a hot torus, and not a thin disc, and moreover that the oscillations are not hindered by the Papaloizou-Pringle instability which rapidly reshapes the torus's angular momentum profile \citep{Papri,Frag05}. The question of what amplifies the modes is also not easily answered. 

Alternatively, one can treat the accretion flow as a \emph{thin disc} extending to the ISCO. HFQPOs can then be associated with the intrinsic oscillations of the thin disc, in particular its inertial waves (here referred to as r-modes). This model is attractive because the hydrodynamic theory predicts that inertial waves should be confined by relativistic effects to an
annular region separated from the ISCO, and consequently take on a global character.\footnote{
Local oscillations are disfavoured because they would generate
a broadband frequency component rather than distinct peaks. See
\citet{DexBl}, however, for a model involving a bandpass
filter.}
Confinement in such a `self-trapping region' would both protect trapped inertial waves from the uncertain (and probably unfavorable) conditions at the ISCO, and also endow the lowest order, fundamental mode with a frequency close to the maximum attained by the horizontal epicyclic frequency, $\kappa$. As a consequence, the wave frequencies would sit directly in the observed range for HFQPOs, possess the correct scaling with black hole mass, and depend on black hole spin in a straightforward way. Finally, and importantly, amplification of these global standing waves to dynamically significant (and observable) levels can be explained via a non-linear coupling with disc warps and eccentricities \citep{Oka87,kat01,Kat04,Kat08,FoG08}. One problem with this model, at least in its linear incarnation, is that it fails to account for multiple HFQPOs. 

This paper adopts the thin disc `diskoseismological' model as a starting point and explores its generalisation to magnetohydrodynamics (MHD). Unsurprisingly, the inclusion of magnetic fields changes the dynamical behavior of trapped inertial modes in thin discs. \citet{FL09} found that the inclusion of a purely constant, vertical magnetic field in a local analysis drives the trapping region toward the ISCO. In fact, the authors suggested that a constant, purely vertical field of sufficient strength, in particular a mid-plane plasma beta (ratio of gas pressure to the magnetic pressure) of $\beta_z\lesssim 300$, would force the inner turning point for the trapped inertial waves to coincide with the inner disc edge. Such a shift in localisation would make r-mode excitation a less attractive explanation for HFQPOs, since the oscillations would then require reflection at the inner boundary, and might be subject to damping by radial inflow \citep{Fer10}. 

In \citet{DLO} (hereafter DLO), we expanded on the local analyses of \citet{FL09}, solving
the 2D eigenvalue problem to compute fully global r-modes in a disc model including both vertical magnetic fields and density stratification. Our results were in rough agreement with \citet{FL09} though we found the severity of the effect was modulated by the vertical structure of the modes and the disc temperature (trapped inertial waves are less well-confined in hotter discs).  Characteristic temperatures of $\sim 1\textrm{keV}$ give estimates of critical plasma betas of $\beta_z\sim 100-300$ below which r-mode trapping relies on reflection at the inner disc edge.

Perhaps of greater significance, DLO noted that a \textit{large-scale},\textit{ net flux} vertical field with mid-plane $\beta_z\lesssim1000$ is in fact significant in that it would strongly modify outflows and turbulence due to the magnetorotational instability (MRI). Such strong magnetic fields may be uncommon, though winds observed in GRS 1915+105 in some emission states have been taken as evidence of magnetic driving \citep[e.g.,][]{Mill16}. In any case, a large-scale, smooth and purely vertical field should be distinguished from the magnetic fluctuations associated with the MRI, which are small-scale, unsteady, and generally stronger. DLO further noted that an inertial wave pushed up against the ISCO might still achieve coherence and observable amplitudes if forced sufficiently strongly by a warp or eccentricity.

In this work we do not argue for or against strongly magnetized discs. Rather, we would like to point out that if a disc were threaded by a strong net vertical field, local simulations, both recent and canonical, indicate that the associated MRI would produce \emph{even stronger toroidal fields}, often approaching equipartition with an associated mid-plane $\beta_\phi \sim 1$ \citep{Haw95,Sto96,Bai13,salv16a}. Similarly, global simulations suggest that the mean toroidal component in an MRI turbulent disc will be much stronger than the mean vertical component \citep[e.g.,][]{Zhu18}. The question then is: does this strong toroidal field have any effect on trapped r-modes? Can it moderate or counter the influence of a large-scale vertical magnetic field? Both \citet{FL09} and DLO examined toroidal fields in isolation and found little different to the hydrodynamical case; the `mixed case' was not treated. The overwhelming numerical and analytical evidence that strong toroidal fields must always accompany strong vertical fields provides motivation for our study.

We find that sufficiently strong toroidal magnetic fields can restore the trapping region's isolation from the ISCO. Concurrently, they reduce the r-modes' magnetically enhanced frequencies toward those predicted by the hydrodynamic theory. A strong toroidal magnetic field component significantly reduces the detrimental effects of a moderate to strong vertical field with mid-plane $\beta_z\lesssim500$, while for larger $\beta_z$ inertial wave trapping remains minimally affected (for any $\beta_{\phi}$). Near-equipartition values of $\beta_{\phi}\gtrsim 1$ may be required to restore the r-mode trapping region when $\beta_z\lesssim200$, but this strong a toroidal component is not unrealistic. We interpret the effect of a toroidal magnetic field as an alteration of the magneto-acoustic response of the trapped inertial modes, which increases the effective plasma beta due to any mean vertical field. Importantly, because the effect arises from an enhancement of the effective pressure, a strong toroidal field need not be ordered on large-scales to impact r-modes, and hence could be supplied by a vigorous MRI. 

We first present our linearized equations and disc model in Section \ref{sec:model}. We then investigate the mixed case of toroidal and vertical background fields through the local linear WKBJ theory in Section \ref{sec:local}, with an analysis similar in spirit to that of \citet{FL09}. In Section \ref{sec:global}, this is generalised to global linear calculations in both cylindrical discs and in fully stratified discs. We provide a discussion of this work's implications and a critical assessment of competing models of HFQPOs in Section \ref{sec:disc}, before concluding in Section \ref{sec:conc}.

\section{Equations and disc model}\label{sec:model}
In this section we introduce our magnetohydrodynamic disc model and present the linearized equations solved throughout the paper.

\subsection{Governing equations}
As in DLO, we presume that the accretion flow in the SPL state can be modelled as a thin, centrifugally supported disc of inviscid, non self-gravitating, ionized gas. Furthermore, it is assumed to extend all the way to the ISCO. Our picture might be compared to that of \citet{Nay0}, but we focus on the disc and defer consideration of a hot corona. The flow can be approximately described by the non-relativistic, ideal MHD equations:
\begin{align}
\label{eq:eqEoM}
    \dfrac{\del {\bf{u}}}{\del t}
    +{\bf u\cdot\nabla u}
    &=-\dfrac{\nabla P}{\rho}
    -\nabla\Phi
    +\dfrac{1}{\mu_0\rho}{\left(\curl\bf{B}\right)\times\bf{B}},
\\[5pt]
\label{eq:Cty}
    \dfrac{\del \rho}{\del t}
    &=-\nabla\cdot \left(\rho{\bf u}\right),
\\[5pt]
\label{eq:IE}
    \dfrac{\del {\bf{B}}}{\del t}
    &=\curl\left({\bf{u\times B}}\right),
\\[5pt]
\label{eq:sol}
    \dv{\bf{B}}&=0.
\end{align}	
Here ${\bf{u}}$, $\rho$, $\bf{B}$, and $P$ are the fluid velocity, density, magnetic field and gas pressure, respectively, and $\Phi$ is an arbitrary gravitational potential. For simplicity and to isolate the effects of magnetic fields, the system is closed by a globally isothermal equation of state, $P=c_s^2\rho$, where $c_s$ is the isothermal sound speed (taken to be constant throughout the disc).

\subsection{Basic, equilibrium state}
In cylindrical coordinates $(r,\phi,z)$, Equations (\ref{eq:eqEoM})-(\ref{eq:sol}) admit an equilibrium of the form ${\bf{u}}=r\Omega(r)\ephi$, in isorotation with a magnetic field ${\bf{B}}=B_{\phi}(r)\ephi+B_z(r)\ez$, the r-component of Equation (\ref{eq:eqEoM}) implying 
\begin{equation}\label{eq:OMmmm}
    \Omega^2(r)
    =\dfrac{1}{r}
    \left[
    	\dfrac{\del\Phi}{\del r}
    	+\dfrac{1}{\rho}\dfrac{\del P}{\del r}
    	+\dfrac{1}{\mu_0\rho}
        \left(
          \dfrac{B_{\phi}}{r}\dfrac{\textrm{d}(rB_{\phi})}{\textrm{d} r}
          +B_z\dfrac{\textrm{d} B_z}{\textrm{d} r}
    	\right)
    \right].
\end{equation}
We take $B_z$ constant and $B_{\phi}\propto r^{-1}$, the latter profile giving the state of minimum magnetic energy for a given azimuthal magnetic flux \citep{Og96}. This configuration conveniently leaves the equilibrium angular velocity profile unaffected by any Lorentz force. However, we have performed vertically local calculations with alternative power laws for $B_{\phi}$, and have found that they do not qualitatively change the results presented here. In short, more negative (positive) gradients in $B_{\phi}$ ($B_z$) increase the robustness of r-mode trapping. 

Since the background magnetic field considered is independent of $z$, it does not contribute to the vertical equilibrium. The assumption of a vertically homogeneous $\bf B$ is an oversimplification, and should certainly be revisited for any investigations aiming to connect dynamics in the disc with a rarefied corona. For our geometrically thin disc, the density profile is written as $\rho=\rho_0(r)g(z/H)$. Here $\rho_0$ is the mid-plane density, $H=c_s/\Omega_z$ is the isothermal scale height (with $\Omega_z=\del_{zz}^2\Phi$ the vertical epicyclic frequency), and $g$ is a dimensionless vertical profile. In the isothermal case $g$ adopts a Gaussian form, $g=\exp[-z^2/(2H^2)]$. 

The effects of a background radial pressure gradient (on both the equilibrium flow and linearized perturbations) were considered in DLO, where it was found that a negative radial power law in $\rho$ marginally increases r-mode resistance to the effects of a vertical magnetic field. Vertically local calculations (excluded from this work for simplicity) suggest that, like alternative radial profiles for the magnetic field, radial gradients in $\rho$ and $c_s$ do not qualitatively change the effect of the strong toroidal magnetic field component considered here. As a result, we set $\rho_0$ equal to a constant in what follows. With this density and magnetic field distribution, Equation \eqref{eq:OMmmm} reduces to $\Omega^2(r)=(1/r)\del_r\Phi$. 

The flow is also unstable to the MRI, and a more realistic model might include some prescription for turbulent damping of large-scale oscillations, and for turbulent heating. We do not believe this physics is essential to understanding the impact of ordered magnetic fields on the geometry of the r-mode trapping cavity, and so accretion and radiation are omitted. Note that the effects of accretion and turbulent damping on trapped inertial waves have been quantified by \citet{Fer10}, who found that a transonic radial inflow introduces r-mode damping and a decay rate that, although substantial for sonic points outside the ISCO, could still be overcome by excitation by a large amplitude warp or eccentricity. We defer an examination of the competition between damping due to radial inflow/turbulence and excitation due to non-linear mode coupling to future work.

\subsection{Characteristic frequencies}
For an approximate description of a relativistic flow around a black hole, $\Phi$ might be taken as a `pseudo-Newtonian' Paczynski-Wiita potential of the form 
\begin{equation}\label{eq:PW}
    \Phi=\dfrac{-GM}{\sqrt{r^2+z^2}-r_S},
\end{equation}
where $r_S=2GM/c^2$ is the Schwarzschild radius, for $c$ the speed of light and $M$ the black hole mass. As reviewed in DLO, the horizontal epicyclic frequency derived from this potential with the Newtonian formula $\kappa^2=2\Omega\left(2\Omega+rd \Omega/d r\right)$ reproduces the non-monotonic behavior apparent in the fully relativistic version of the epicyclic frequency \citep{Oka87}. 

In the linear theory, however, our arbitrary potential $\Phi$ disappears entirely from the perturbed equations, leaving only the characteristic frequencies $\kappa$ and $\Omega$. A common practice is therefore to utilize the fully relativistic versions of the characteristic frequencies for particles in orbit around a Kerr black hole in an otherwise hydrodynamical or magnetohydrodynamical treatment. This then permits the approximate inclusion of black hole spin into the problem, without requiring a fully general relativistic, magnetohydrodynamic treatment. In units of $r_g=GM/c^2$ and $\omega_g=c^3/(GM)$, these expressions are given by 
\begin{align}
\label{eq:KerrOm1}
\Omega_G
&=\dfrac{1}{(r^{3/2}+a)},\\[8pt]
\label{eq:KerrOm2}
\kappa_G\ 
&=\Omega_G\sqrt{1-\dfrac{6}{r}+\dfrac{8a}{r^{3/2}}-\dfrac{3a^2}{r^2}},\\[8pt]
\label{eq:KerrOm3}
\Omega_{Gz}
&=\Omega_G\sqrt{1-\dfrac{4a}{r^{3/2}}+\dfrac{3a^2}{r^2}},
\end{align}
where $a\in(-1,1)$ is the dimensionless spin angular momentum parameter. 

Since Equations (\ref{eq:KerrOm1})-(\ref{eq:KerrOm3}) describe the orbits of particles, even in linear theory they would be inconsistent with a background fluid flow modified by a strong pressure gradient or Lorentz force. However, the fully relativistic versions of the characteristic frequencies are appropriate for use with a background state such as the one considered here (i.e.\ constant $\rho_0$ and $B_z$, $B_\phi\propto 1/r$), and allow for inclusion of the effects of black hole spin. In addition, Equations (\ref{eq:KerrOm1})-(\ref{eq:KerrOm3}) have the advantages of correctly reproducing the radius of marginal stability, the radius of maximal $\kappa$, and the rates of nodal and apsidal relativistic precession.

\subsection{Linearized equations}
Axisymmetric r-modes are of the most physical and observable interest. Non-axisymmetric r-modes are strongly damped at their corotation radii where the mode frequency $\omega=m\Omega$, with $m$ the azimuthal mode number \citep{Li02}. While some modes might avoid this damping if their frequencies are so large that the corotation radius lies outside of the trapping region, such frequencies would be too high for measured HFQPOs, even for small $m=1,2,...$ and low values of the spin angular momentum parameter \citep{Wag12}. Further, even if non-axisymmetric modes are an essential component of the excitation mechanism considered by \citet{FoG08}, any observable signature is likely to be provided by the fundamental, axisymmetric r-mode with simplest radial and vertical structure. 

For this reason, we consider axisymmetric, Eulerian perturbations of the form $\operatorname{Re}\{\delta (r,z)\exp[\text{i}\omega t]\}$ to the equilibrium state. Linearizing Equations (\ref{eq:eqEoM})-(\ref{eq:sol}) and making use of the solenoidal condition yields
\begin{align}
\label{eq:lowE1}\notag 
    -\ti\omega v_r\ 
    &=2\Omega v_{\phi}
    -\dfrac{\del h}{\del r}
    \\&+\dfrac{1}{g}
    \left(
    	V_{\tA z}\dfrac{\del v_{Ar}}{\del z}
    	-\dfrac{V_{\tA\phi}}{r}\dfrac{\del (rv_{\tA\phi})}{\del r}
        -V_{\tA z}\dfrac{\del v_{\tA z}}{\del r}
    \right),
\\[5pt]
\label{lowE2}
    -\ti\omega v_{\phi}\ 
    &=-\dfrac{\kappa^2}{2\Omega}v_r 
    +\dfrac{V_{\tA z}}{g}\dfrac{\del v_{\tA\phi}}{\del z},
\\[5pt]
\label{lowE3}
    -i\omega v_z\ 
    &=-\dfrac{\del h}{\del z}
    -\dfrac{V_{\tA\phi}}{g}\dfrac{\del v_{\tA\phi}}{\del z},
\\[5pt]
    -\ti\omega h\ \ 
    &=-c_s^2\left(
    	\dfrac{1}{rg}\dfrac{\del (rg v_r)}{\del r}
        +\dfrac{1}{g}\dfrac{\del (gv_z)}{\del z}
	\right),
\\[5pt]
    -i\tomega v_{Ar}
    &=V_{\tA z}\dfrac{\del v_r}{\del z},
\\[5pt]
\notag 
    -\ti\omega v_{\tA\phi}
    &=-V_{\tA \phi}\dfrac{\del}{\del \ln r}\left(\dfrac{v_r}{r}\right)
    \\
\label{eq:lowE6}
    &\hspace{1cm}
    +V_{\tA z}\dfrac{\del v_{\phi}}{\del z}
    -V_{\tA\phi}\dfrac{\del v_z}{\del z}
    +\dfrac{\td\Omega}{\td \ln r}v_{Ar},
\\[5pt]
\label{eq:lowE7}
    -\ti\omega v_{\tA z}
    &=-\dfrac{V_{\tA z}}{r}\dfrac{\del (rv_r)}{\del r},
\end{align}
where ${\bf v}$ is the velocity perturbation, $h=\delta P/\rho$ is the enthalpy perturbation, ${\bf V}_{\tA}={\bf B}/\sqrt{\mu_0\rho_0}$ is  the mid-plane Alfv\'en velocity of the background magnetic field, and ${\bf v}_{\tA}=\delta {\bf B}/\sqrt{\mu_0\rho_0}$. We use Equations (\ref{eq:lowE1})-(\ref{eq:lowE7}) to derive a local dispersion relation in Section \ref{sec:local}, calculate radially global but vertically local normal modes in Section \ref{sec:cglobal} and solve for fully global solutions in Section \ref{sec:fglobal}.

\section{Local dispersion relation}\label{sec:local}
We begin by conducting a local analysis of wave propagation in the disc model outlined in Section \ref{sec:model}. Local approximations are strictly inappropriate for the description of r-modes global in nature. They do, however, provide qualitative insight into why toroidal magnetic fields have a larger impact on trapped inertial waves when considered in combination with a poloidal field component, rather than in isolation. 

We assume that the perturbations possess radial and vertical wavelengths much smaller than the scale of variation for the background flow, and thus prescribe the dependence $\delta(r,z)\propto\exp[\ti k_rr+\ti k_zz]$. Here $k_r$ and $k_z$ are assumed both to be $\gg 1/r$ and slowly varying with radius, such that their radial derivatives may be neglected. We also concentrate on a small region at a fixed radius $r$ in the disc. Vertical variation in $\rho$, and terms $\propto 1/r$ are hence neglected as sub-dominant in this approximation, although $\kappa$, $\Omega$, $r\del_r\Omega$ and $V_{\tA\phi}$ may be regarded as functions of the fixed radius, varying as we examine mode behavior at separate locations.

With these assumptions, equations (\ref{eq:lowE1})-(\ref{eq:lowE7}) can be reduced to a bi-cubic dispersion relation
\begin{align}\label{eq:disp1}
    &\omega ^6
    -\left[
          \kappa^2
        + k^2\left(c_s^2 +V_{\tA}^2\right)
        + k_z^2 V_{\tA z}^2
    \right]\omega ^4
\\\notag 
&\hspace{.5cm}
    +k_z^2\left[
        \kappa^2 \left(c_s^2 + V_{\tA\phi}^2\right)
        + V_{\tA z}^2 
        \left(
              k^2 \left[2 c_s^2+V_{\tA}^2\right]
            +\dfrac{\td\Omega^2}{\td\ln r}
        \right)
    \right]\omega ^2 
\\\notag 
&\hspace{4cm}
    -c_s^2 k_z^4 V_{\tA z}^2
    \left(
          k^2 V_{\tA z}^2
        + \dfrac{\td\Omega^2}{\td\ln r}
    \right)=0,
\end{align}
where $k^2=k_r^2+k_z^2$ and $V_{\tA}^2=V_{\tA \phi}^2+V_{\tA z}^2$. Equation (\ref{eq:disp1}) offers immediate insight into the physical nature of the effect that a background toroidal magnetic field might have on trapped inertial waves, since $V_{\tA\phi}^2$ appears only in sum with $c_s^2$. The work of the background azimuthal field here may be understood to \emph{increase the effective sound speed}, which in turn increases the effective plasma beta associated with the vertical magnetic field (therefore reducing its detrimental effect).

Assessing the trapping of inertial waves through a local method reduces to a study of the local radial wavenumber, as the regions of the disc in which the perturbations may be expected to be oscillatory are those in which $k_r$ is real, or $k_r^2>0$. The radial profile of $-k_r^2$ can be thought of as an effective potential well, implying oscillatory (evanescent) behavior wherever $-k_r^2<0$ $(-k_r^2>0)$. The vertical wavenumber, on the other hand, may be considered a free parameter. However, as shown in DLO, prescribing the vertical wavenumber $k_z=K_n/H$ accurately reproduces dynamics of the global r-mode spectrum, where $K_n$ are dimensionless eigenvalues associated with the basis functions of order $n$ describing r-modes' vertical structure in the presence of a purely vertical magnetic field. We revisit this prescription in section \ref{sec:fglobal}, but find that it remains reasonably accurate when toroidal fields are included.

It is useful (and physically intuitive) to define the characteristic frequencies $\omega_{\tA z}=k_zV_{\tA z}$ and $\omega_{\tA \phi}=k_zV_{\tA\phi}$. Then, solving for $k_r^2$ from equation (\ref{eq:disp1}) yields after some algebra
\begin{equation}\label{eq:disp2}
k_r^2
=\dfrac{
(\omega^2-\omega_{\textrm{FL1}}^2)
(\omega^2-\omega_{\textrm{FL2}}^2)
(\omega^2-\omega_{\textrm{FL5}}^2)
-\omega_{\tA\phi}^2(\omega^2-\omega_{\kappa B}^2)\omega^2
}
{
  (V_{\tA}^2+c_s^2)
  \left(\omega^2-\omega_{\textrm{FL3}}^2\right)
  \left(\omega^2-\omega_{\textrm{FL4}}^2\right)
},
\end{equation}
where
\begin{align}
    \omega_{\textrm{FL1}}^2
    &=k_z^2c_s^2,
\\\label{eq:omFL2}
    \omega_{\textrm{FL2}}^2
    &=\dfrac{1}{2}
    \left[
       \kappa^2 
       +2\omega_{\tA z}^2
       +\sqrt{
            \kappa^4 
            +4\omega_{\tA z}^2
            \left(
                \kappa^2
          	    -\dfrac{\td\Omega^2}{\td\ln r}
          	\right)
       }
    \right],
\\
    \omega_{\textrm{FL3}}^2
    &=\omega_{\tA z}^2,
\\  
    \omega_{\textrm{FL4}}^2
    &=\dfrac{\omega_{\tA z}^2c_s^2}{V_{\tA}^2+c_s^2},
\\
    \omega_{\textrm{FL5}}^2
    &=\dfrac{1}{2}
    \left[
       \kappa^2 
       +2\omega_{\tA z}^2
       -\sqrt{
            \kappa^4 
            +4\omega_{\tA z}^2
            \left(
                \kappa^2
          	    -\dfrac{\td\Omega^2}{\td\ln r}  
          	\right)
       }
    \right],
\end{align}
are the characteristic frequencies identified by \citet{FL09} in deriving a dispersion relation for a purely constant, vertical field (cf. their equations 30-35), and we have defined
\begin{equation}
	\omega_{\kappa B}^2
    =\kappa^2 + \omega_{\tA z}^2.
\end{equation} 
For $B_{\phi}=0$ or $B_z=0$, equation (\ref{eq:disp2}) reduces to \citet{FL09}'s equations (30) and (41), respectively, although we have ignored the gradient terms appearing in the latter because of our assumption $k_r,k_z\gg 1/r$. 

\subsection{Trapping region}
We now discuss the nature of r-mode trapping. At a given radius, the roots of Equation (\ref{eq:disp2}) separate regions in frequency space that exhibit either oscillatory or evanescent behavior based on whether $k_r^2>0$ or $k_r^2<0$ (resp.), with resonance at $k_r=0$. Solving for these resonant frequencies at all radii then provides the structure of oscillatory behavior in the disc. If one root takes the same value of $\omega^2$ at two nearby radii, in such a way that $k_r^2>0$ in between the radii but $k_r^2<0$ outside, we might expect the radii to define turning points for a trapped global oscillation. 

In a purely hydrodynamic disc ($V_{\tA}=\omega_{\tA z}=\omega_{\tA\phi}=0$), inertial waves being both `slow', and animated by the restoring force of rotation, possess frequencies less than the local epicyclic frequency, $\kappa$. The turning points then delimit a region in which $\omega<\kappa$. Because $\kappa$ varies with radius non-monotonically, this trapping region is well-defined and separate from the inner boundary. This is shown by the dashed curve in Fig.~\ref{fig:trapcalc}, which describes $\kappa(r)$. An illustrative hydrodynamic r-mode is superimposed as the squiggly line. 

Now let us include a purely vertical field ($\omega_{Az}\neq 0,
\omega_{A\phi}=0$). \citet{FL09} showed that the r-mode resonances and consequent trapping regions are modified, and this is clear from the numerator in
\eqref{eq:disp2}. The natural frequency upon which resonance occurs (i.e.\ $k_r=0$)
increases from $\kappa$ to $\omega_\text{FL2}$, which can be significantly larger because of magnetic tension (see Equation \ref{eq:omFL2}). Now when $\omega=\omega_\text{FL2}$, the inertial wave comes not into resonance with a hydrodynamic epicycle but a rotationally modified Alfv\'en wave, propagating vertically. An important consequence of this increase is that the inner turning point for an r-mode of a given frequency shifts inward, toward the ISCO. This is illustrated by the black curves in both panels of Fig.~\ref{fig:trapcalc}, which describe the frequency $\omega_\text{FL2}(r)$ for two different treatments of vertical structure. Under certain conditions, in particular when the field is very strong, $\omega_{\textrm{FL2}}^2$ can lose a maximum distinct from the ISCO. The inner disc edge must then provide an inner reflection point if the r-modes are to remain confined (see bottom panel in Fig. \ref{fig:trapcalc}).

Adding an azimuthal field further alters the resonances, via the last
term in the numerator of Eq.~(\ref{eq:disp2}). Its effect is
complicated and not easy to distentangle. What is clear is that the
azimuthal field introduces a coupling between the epicyclic and
Alfvenic response of the disc, on one hand, and its acoustic response,
on the other (see Section \ref{sec:asymp}).
The latter is absent when there is only a vertical field, as can be
seen by the absence of $c_s$ in the expression for
$\omega_\text{FL2}$. We shall see, in the following sections, that
this acoustic coupling decreases the resonant frequency, moving the
trapping region away from the ISCO and back towards its hydrodynamical
location.

\subsection{Asymptotic analysis for a weak poloidal
  field}\label{sec:asymp}

We consider the limit in which the vertical magnetic field
component is assumed weak (i.e., $\beta_z\gg 1$), while the toroidal
field is allowed to take any value. Expanding the frequency as
$\omega=\omega_0+\beta_z^{-1}\omega_1+\mathcal{O}(\beta_z^{-2})$,
 the dispersion relation reduces at zero'th order in $\beta_z^{-1}$ to
\begin{equation}
	k_r^2=
    \dfrac{
    	\left[\omega_0^2-k_z^2\left(c_s^2+V_{\tA\phi}^2\right)\right]
    	(\omega_0^2-\kappa^2)
    }{\omega_0^2\left(c_s^2+V_{\tA\phi}^2\right)}.	
\end{equation}
This dispersion relation is nearly identical in form to the
hydrodynamic dispersion relation for axisymmetric modes (cf. equation 1 in DLO). It describes magneto-acoustic
oscillations that propagate where
$\omega^2>k_z^2(c_s^2+V_{\tA\phi}^2)$, and r-modes, which are
unaffected by the toroidal magnetic field and remain trapped where
$\omega^2<\kappa^2$.

Dispensing with the magneto-acoustic oscillation, we choose one of the r-modes by setting $\omega_0= \kappa$ and compute the next
order correction due to the magnetic field. In summary, we find
\begin{equation}\label{eq:ord1exp}
    \omega^2=\kappa^2
    +\omega_{\tA z}^2
    \left[  
        1+\left(\dfrac{4\Omega^2}{\kappa^2}\right)
        \left(
            \dfrac{k_z^2 c_s^2-\kappa^2}
            {k_z^2\left(c_s^2+V_{\tA\phi}^2\right)-\kappa^2}
        \right)
    \right]
    +\dots,
\end{equation}
where we have used the expression $\kappa^2=2\Omega\left(2\Omega+rd \Omega/d r\right),$ which is only approximate for characteristic frequencies defined by Equations (\ref{eq:KerrOm1})-(\ref{eq:KerrOm3}).

A number of things can be said about Eq.~\eqref{eq:ord1exp}. First, the magnetic correction to the frequency is always positive. This can be proven by noting that the vertically local approximation requires $k_z>1/H=\Omega_z/c_s$, and that in a relativistic disc $\Omega_z^2>\kappa^2$. Thus $k_z^2 c_s^2 - \kappa^2 > \Omega_z^2-\kappa^2 > 0$, and the numerator is positive. It follows that the denominator is also positive.

Second, though the magnetic correction enhances the resonant frequency, increasing the azimuthal field $V_{A\phi}$ reduces this effect, as it only appears in the denominator. Moreover, the mid-plane azimuthal Alfv\'en speed only appears squared in a sum with $c_s^2$, i.e. as an intensification of the effective pressure in the disc (as pointed out earlier by direct inspection of the dispersion relation \ref{eq:disp1}). It is clear that the azimuthal field couples the thermal response of the disc to the r-mode oscillations, providing an additional effective acoustic response. In so doing it reduces the resonant frequency defining the r-mode trapping region.

We find that a strong toroidal magnetic field component results in an altered \textit{effective} plasma beta of the vertical magnetic field, $\beta_{z,e}>\beta_z$. Heuristically, for $\beta_{\phi}\gtrsim1$, implementing a purely vertical field with 
\begin{equation}
    \beta_{z,e}=\beta_z(1+2/\beta_{\phi})
\end{equation}
produces similar results (in both local and global calculations) to the corresponding mixed field. 

\subsection{General trapping region calculations}\label{sec:numdisp}
\begin{figure}                           
        \includegraphics[width=.5\textwidth]{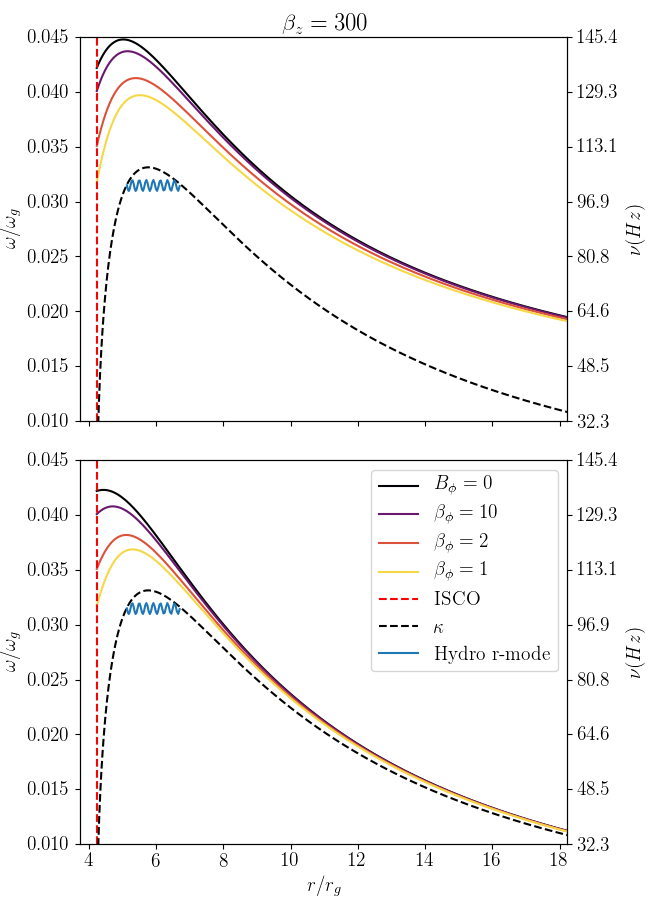}    
\caption{
Numerically calculated r-mode trapping regions for $a=0.5$, $\beta_z=300$ and increasing values of azimuthal magnetic field strength (decreasing $\beta_{\phi}$). The top (bottom) plot shows calculations with $k_z=K_1/H(r_{\textrm{ISCO}})$ ($k_z=K_1/H(r)$), where $K_1\approx1.158$ (see Section \ref{sec:cglobal}). The red and black dashed lines mark the ISCO and the hydrodynamic trapping region, respectively. Angular frequencies $\omega$ in $\omega_g=c^3/(GM)$ are given on the left, while the corresponding frequencies $\nu$ in Hz calculated with the assumption $M_{BH}=10M_{\odot}$ are given on the right.}
\label{fig:trapcalc}
\end{figure}

A brute force approach to confirming the asymptotic analyses of Section \ref{sec:asymp} is to solve the bi-cubic appearing in the numerator of Equation (\ref{eq:disp2}) numerically at each radius. Doing so provides radial profiles for the resonant frequencies described in Section \ref{sec:local}, one of which can be easily identified as defining the r-mode trapping region. Trapping regions calculated in this way are plotted with colored solid lines in Fig. \ref{fig:trapcalc} for $a=0.5$, a fixed value of $\beta_z=300$ and increasing azimuthal field strength. 

Fig. \ref{fig:trapcalc} (top) shows profiles calculated with $k_z\propto 1/H$ and the approximation of a scale height $H=H(r_{\textrm{ISCO}})$ constant with radius, and Fig. \ref{fig:trapcalc} (bottom) shows calculations made with $H=H(r)$. Assuming a constant $H=H(r_{\textrm{ISCO}})$ is inconsistent with a globally isothermal equation of state, while a scale height $H(r)=c_s/\Omega_z(r)$ increases more rapidly with radius than might be expected in the radiation-pressure dominated inner regions of a black hole accretion disc. We provide calculations made with both approximations for comparison, positing that the two cases may bracket reality. For $B_{\phi}=0$ the local estimate of the trapping region is defined by $\omega_{\textrm{FL2}}$, but increasing the toroidal field strength drives the r-mode trapping region back toward the hydrodynamic profile given by $\kappa$. In other words, a toroidal field reverses the effect of the vertical field.

\begin{figure*}
\centering
\subfigure[$k_z\propto 1/H(r_{\textrm{ISCO}})$]
{
	\includegraphics[width=.48\textwidth]{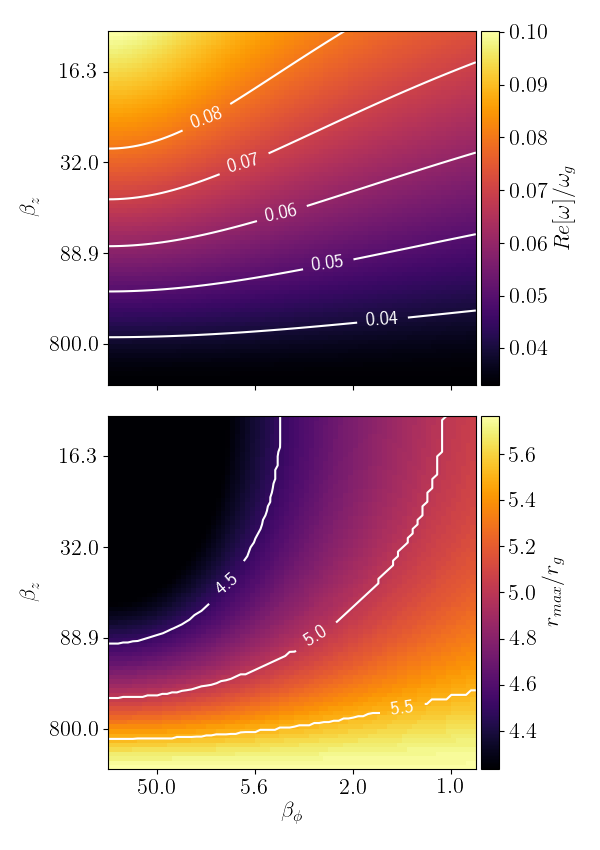}
}
\subfigure[$k_z\propto 1/H(r)$]
{
	\includegraphics[width=.48\textwidth]{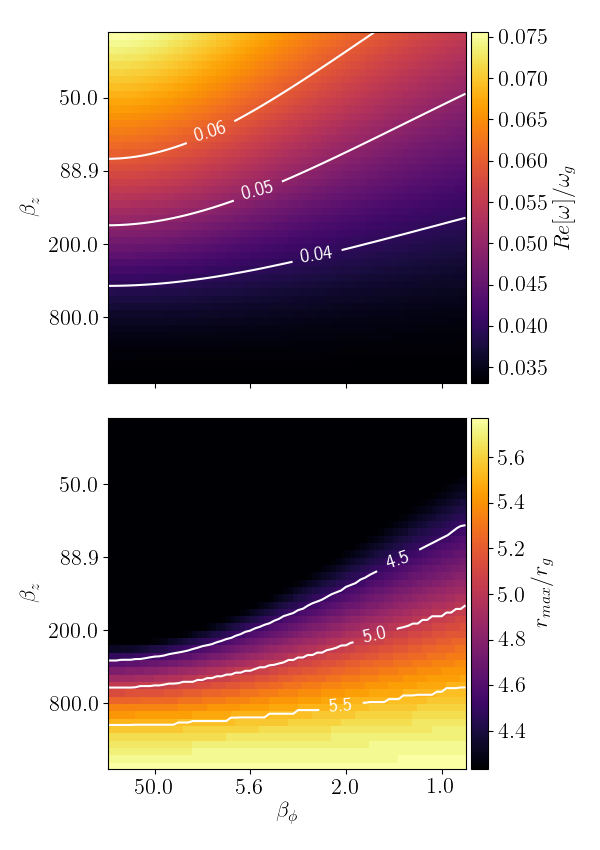}
	}
    \caption{Heatmaps showing the maximal frequencies and radii at which they are obtained for radial profiles of the resonant WKBJ frequency (as illustrated in Fig. \ref{fig:trapcalc}), calculated with $a=0.5$, varying vertical ($\beta_z$) and azimuthal ($\beta_{\phi}$) field strengths, and both $k_z=K_1/H(r_{\textrm{ISCO}})$ (left) and $k_z=K_1/H(r)$ (right). These quantities can be associated with the frequencies and regions of localisation for the fundamental r-mode with the simplest radial structure (resp.). For reference, the ISCO is located at $r\approx 4.233r_g$ for $a=0.5$.}
\label{fig:loc_omsmx}
\end{figure*}

Figs. \ref{fig:loc_omsmx} (top) give heatmaps of the maximal frequencies associated with radial profiles calculated as in Fig. \ref{fig:trapcalc} with increasing vertical and azimuthal field strengths, while Figs. \ref{fig:loc_omsmx} (bottom) show the radii at which this maximum is achieved, denoted as $r_{\max}$. The maximal frequency provides an estimate of the frequency of the fundamental r-mode with the simplest radial structure, and the radii at which it is achieved predicts this mode's region of localisation. Once $r_{\max}\sim r_{\textrm{ISCO}}$ ($r_{\textrm{ISCO}}\approx 4.233r_g$ for $a=0.5$), the inner turning point has been eliminated and trapping isolated from the ISCO is no longer possible. However, along with Fig. \ref{fig:trapcalc}, Figs. \ref{fig:loc_omsmx} indicate that the isolation of the trapping region from the inner disc edge is restored as larger and larger toroidal magnetic field strengths are introduced.

Figs. \ref{fig:trapcalc} and \ref{fig:loc_omsmx} confirm the predictions of our asymptotic analyses, suggesting that a sufficiently strong toroidal field reduces the frequency enhancement provided by a net vertical field, and may even restore the independence of the r-mode trapping region from the inner disc edge. Trapping regions calculated with radial scale height variation show a similar reduction in maximal frequency, but a less drastic increase in maximal radius for large $B_z$. However, as indicated in Figs. \ref{fig:trapcalc} (bottom), for values of $\beta_z\sim 100-300$ that are actually rather strong for a large-scale ordered vertical magnetic field, an azimuthal magnetic field component of equipartition strength still restores the isolation of the trapping region from the ISCO.  

\section{Global calculations}\label{sec:global}
In this section we explicitly calculate global r-mode solutions, confirming the predictions made through local analyses in Section \ref{sec:local} that strong toroidal magnetic fields reduce the frequency enhancement and inward forcing caused by purely vertical fields. Vertically local but radially global calculations are discussed in Section \ref{sec:cglobal}, and fully global calculations in Section \ref{sec:fglobal}.
\subsection{Cylindrical calculations}\label{sec:cglobal}
In the cylindrical model, density stratification and vertical gravity
are ignored (i.e., $g=1\Longrightarrow\rho=\rho_0$) with the
application of a vertically local approximation and the assumption
that axisymmetric perturbations have the dependence
$\delta(r,z,t)\propto\tilde{\delta}(r)\exp[\ti k_zz-\ti \omega
t]$. This model, a radial analogue of the stratified shearing box,
focuses on the mid-plane of the disc, and is attractive from a
numerical standpoint. A continuum spectrum of modes in $k_z$ does
introduce ambiguity, and misrepresents the discrete spectrum uncovered
with fully global calculations in DLO. 
However, as mentioned in Section \ref{sec:local}, with a particular choice of $k_z=K_1/H$, where $K_1\approx 1.158$ (see the Appendix and DLO), the cylindrical model very closely reproduces the dynamical features of the fundamental r-mode calculated in a model including density stratification.
 
Within this simplified framework, derivatives with respect to $z$ in Equations (\ref{eq:lowE1})-(\ref{eq:lowE7}) are replaced by $\ti k_z$, and the system reduces from a set of partial differential equations to a set of ordinary differential equations. Scaling time by $\omega_g^{-1}$, lengths by $r_g$ and velocities by $c$, these ODEs can be re-formulated as a generalized eigenvalue problem for the frequency $\omega$, which we solve using a Chebyshev pseudo-spectral method \citep{boyd}. 

The system is of second order and so requires two boundary conditions, one at each radial boundary. With the choice of a constant $k_z=K_1/H(r_{\textrm{ISCO}})$ (consistent with the cylindrical model), we find that the potential barrier separating r-modes and the outer disc is very large, and wave leakage negligible. We therefore implement a purely rigid outer boundary condition in this case. We also consider a radially varying $k_z=K_1/H(r)$ for comparison with previous calculations, ignoring the coupling of vertical modes that comes from radial variation in the scale height $H$. In this case we implement a wave propagation boundary condition and impose $\del_rv_r=\ti k_rv_r$ at the outer radius, where $k_r$ is determined using Equation (\ref{eq:disp2}) with $\omega$ taken as the maximal frequency from a trapping region calculated as in Section \ref{sec:numdisp}. 

For well-confined trapped inertial modes, the inner boundary condition also makes no difference. However, sufficiently strong vertical magnetic fields cause the r-modes to rely on reflection at the inner boundary. In DLO, we (somewhat arbitrarily) identified this critical field strength as that at which frequencies deviated by $0.01$ per cent (the level of accuracy allowed by our numerical technique) for modes calculated with the inner boundary conditions $v_r=0$ versus $\del_r\delta B_r=0$, and we do the same here. Motivated by \citet{Ker04}, we also consider an inner boundary condition in which the total pressure $\Pi = P+B^2/(2\mu_0)$ is held constant (i.e., $\delta \Pi=0$). This boundary condition results in a slightly more pronounced effect of $B_{\phi}$ on localisation at high $B_z$ than the more neutral condition $\del_r\delta B_r=0$, but provides very similar estimates of critical vertical magnetic strengths as determined by frequency divergence.

\begin{figure}                           
        \includegraphics[width=.49\textwidth]{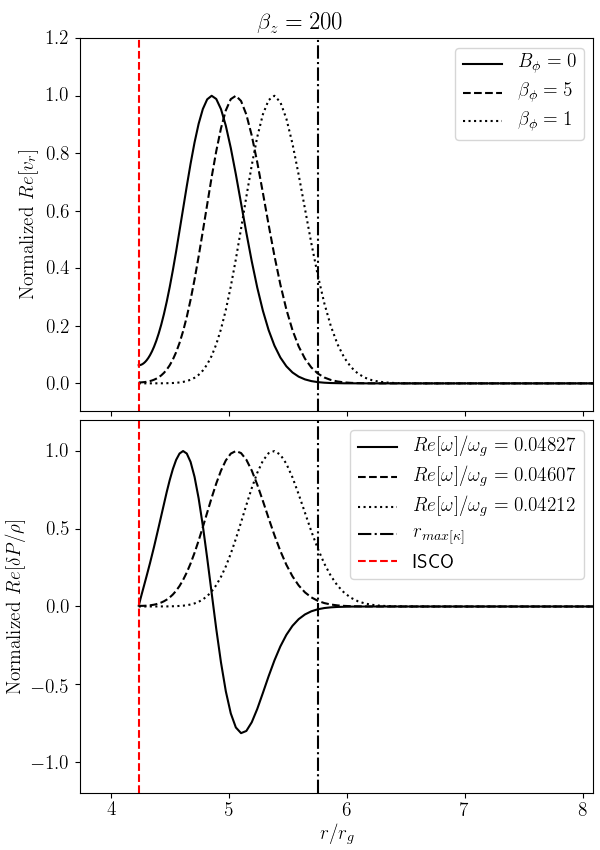}
        \caption{Real parts of the radial velocity (top) and enthalpy (bottom) perturbations for r-modes calculated in the cylindrical model with constant $c_s=0.003c$, $a=0.5$, $k_z=K_1/H(r_{\textrm{ISCO}})$, $\beta_z= 200$, the inner boundary condition $\del_r\delta B_r=0$, and increasing values of the azimuthal magnetic field (the black dash-dotted line shows the radius of maximal $\kappa$). With increasing $B_{\phi}$ the radial quantum number of the enthalpy perturbation changes from $l=1$ to $l=0$ and grows in relative amplitude (not shown), indicating an increasingly compressible mode.}
        \label{fig:cyl_ur_hh}
\end{figure}

\subsubsection{Results}
Fig. \ref{fig:cyl_ur_hh} shows representative radial profiles of the radial velocity (top) and enthalpy (bottom) perturbation for r-modes calculated with a  vertical field providing a mid-plane plasma beta of $\beta_z=200$ and increasingly stronger azimuthal magnetic field strengths. As shown, a strong $B_{\phi}$ counteracts the effects of the vertical field, forcing the fundamental r-mode outward into the disc, and lowering its frequency back toward hydrodynamic values. Additionally, increasing azimuthal field strength alters the compressible nature of the r-modes; As shown in Fig. \ref{fig:cyl_ur_hh} (bottom), the enthalpy perturbation (as well as $v_z$) goes from having a radial quantum number of $l=1$ to $l=0$ for near equipartition $B_{\phi}$.

\begin{figure*}
\centering
\subfigure[$k_z\propto 1/H(r_{\textrm{ISCO}})$]
{
	\includegraphics[width=.48\textwidth]{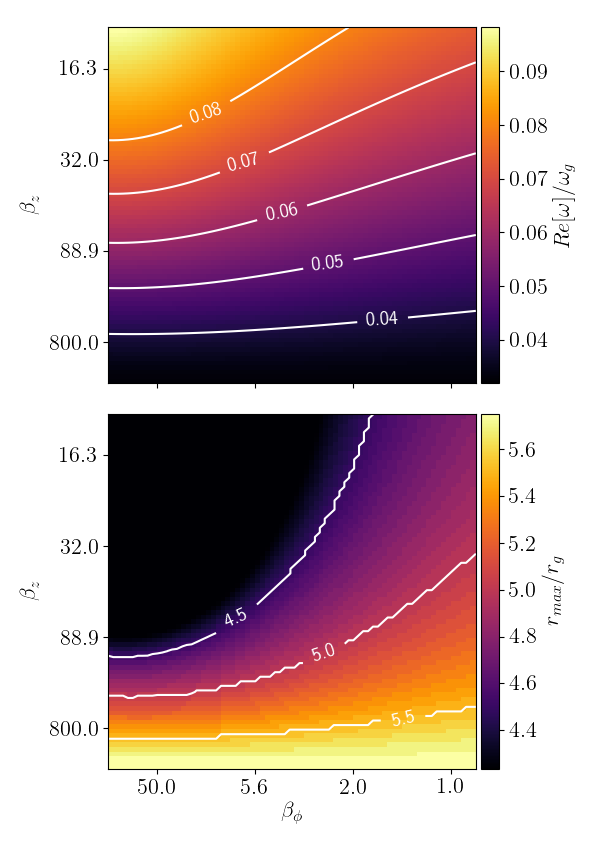}
}
\subfigure[$k_z\propto 1/H(r)$]
{
	\includegraphics[width=.48\textwidth]{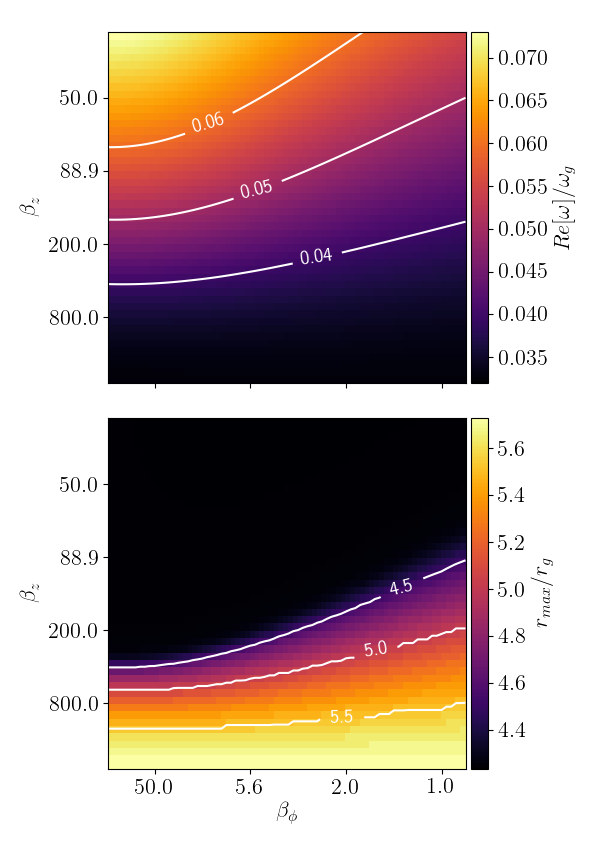}
}
\caption{Radially global calculations analogous to the local ones given in Figs. \ref{fig:loc_omsmx}: Frequencies calculated for the fundamental r-mode with varying $\beta_z$ and $\beta_{\phi}$ are given in the top plots, while the radii at which the modes' radial velocity perturbation achieves its maximum, $r_{\max}$, are given on the bottom ($c_s/c=0.005c$, $a=0.5$, $k_z=K_1/H(r_{\textrm{ISCO}})$ on left, $k_z=K_1/H(r)$ on right, inner boundary condition $\del_r\delta B_r=0$). For reference, the ISCO is located at $r\approx 4.233r_g$ for $a=0.5$.}
\label{fig:cyl_omsmx}
\end{figure*}

Fig. \ref{fig:cyl_omsmx} shows heatmaps of the frequencies $\omega$ (top) and the maximal radii $r_{\max}$ (bottom) of the radial velocity perturbation for the fundamental r-mode, calculated using the cylindrical model with the prescriptions $k_z=K_1/H(r_{\textrm{ISCO}})$ (left) and $k_z=K_1/H(r)$ (right), $c_s=0.005c$, $a=0.5$ and the inner boundary condition $\del_r\delta B_r=0$. The calculated mode frequencies are close to the frequencies predicted by our calculations of the trapping region from the local dispersion relation regardless of inner BC (cf.\ the local calculations given in Fig. \ref{fig:loc_omsmx}), although when \textit{both} $B_{\phi}$ is near equipartition and $B_z$ is strong they are nominally larger than the WKBJ predictions. The global r-modes' locations can vary with the choice of inner boundary condition, however. In particular, for the inner boundary condition $\delta \Pi=0$, $r_{\max}$ separates from $r_{\textrm{ISCO}}$ at lower $B_{\phi}$ than predicted by local calculations, while a stronger $B_{\phi}$ is required for the condition $\del_r\delta B_r=0$.

\begin{figure}                           
        \includegraphics[width=.5\textwidth]{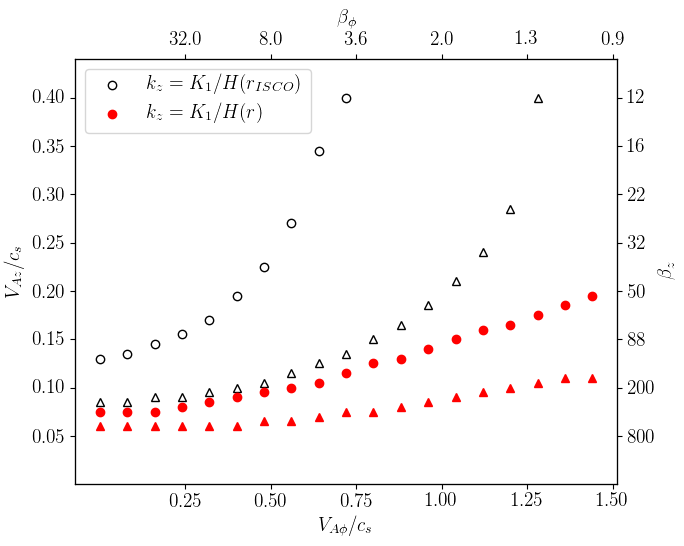}
        \caption{Plots of critical mid-plane vertical magnetic field strengths (y-axis), as a function of azimuthal magnetic field strength (x-axis) found using cylindrical normal mode calculations. The dots indicate the values of $\beta_z$ (right) or $V_{\tA z}/c_s$ (left) at which the inner turning point where $k_r^2(\omega)=0$ coalesces with the ISCO, while triangles show the critical vertical field strengths at which frequencies of r-modes calculated with different inner boundary conditions diverge by $0.01$ per cent ($c_s=0.005c,$ $a=0.5$, $k_z=K_1/H(r_{\textrm{ISCO}})$ for black,  $K_1/H(r)$ for red).}
        \label{fig:cyl_bzcrit}
\end{figure}

Fig. \ref{fig:cyl_bzcrit} shows, for different $k_z$ prescriptions and at a given azimuthal magnetic field strength, the critical strengths of the vertical field at which inertial wave trapping begins to rely on reflection at the inner boundary. Points in the parameter plane that fall above (below) a given curve yield r-modes that are pushed up against (isolated from) the ISCO. As in DLO, we quantify one type of critical field strength curve (marked with triangles) by fixing $\beta_{\phi}$ and finding the value of $\beta_z$ at which r-mode frequencies disagree for different inner boundary conditions. We also include critical curves determined by equating the inner turning point (set by $k_r^2=0$) with the ISCO (marked with dots). For both metrics, the estimates of critical magnetic field strength are largely independent of which inner boundary conditions are used. Fig. \ref{fig:cyl_bzcrit} shows that for a given $\beta_z$, increasing azimuthal field strength and moving from left to right can result in passage from a regime in which r-mode trapping requires reflection at the ISCO to a regime in which the trapping is independent.

Varying the spin parameter $a$ changes frequency and localization only as it might in hydrodynamic calculations, by modifying the relativistic versions of the characteristic frequencies, and the location of the ISCO. The sound speed has little impact on the frequency or localization. However, as discussed in DLO, a larger value of $c_s$ widens the effective potential well. With the inclusion of radial variation in $H$, this results in modest decay rates for sound speeds $\gtrsim 0.005c$ that are larger in amplitude with larger $B_z$. Larger values of $c_s$ also make interaction with the inner disc edge possible at lower vertical magnetic field strengths. 

\subsection{Fully global calculations}\label{sec:fglobal}
In this section we present fully global calculations of axisymmetric r-modes, with the goal of validating the choice of vertical wavenumber $k_z=K_1/H$ made for the local and radially global analyses presented in Sections $3$ and $4.1$. It is convenient to trade the enthalpy perturbation for $\Gamma=\delta\rho/\rho$ and the variables $v_{\tA \phi}$ and $v_{\tA z}$ for
\begin{align}
    \Lambda 
    &=V_{\tA z}\dfrac{\del v_{\tA r}  }{\del z }
    -\dfrac{V_{\tA\phi}}{r}\dfrac{\del (rv_{\tA\phi})}{\del r}
    -V_{\tA z}\dfrac{\del v_{\tA z}}{\del r},
\\[10pt]
    \Theta 
    &=\dfrac{\del v_{\tA\phi}}{\del z}.
\end{align}
$\Lambda$ is proportional to the radial and $\Theta$ to both the azimuthal and vertical components of the Lorentz force perturbation. The evolutionary equations for these variables (Equations \ref{eq:strat6} and \ref{eq:strat7}) then replace Equations (\ref{eq:lowE6}) and (\ref{eq:lowE7}).

\begin{figure*}                           
        \includegraphics[width=.95\textwidth]{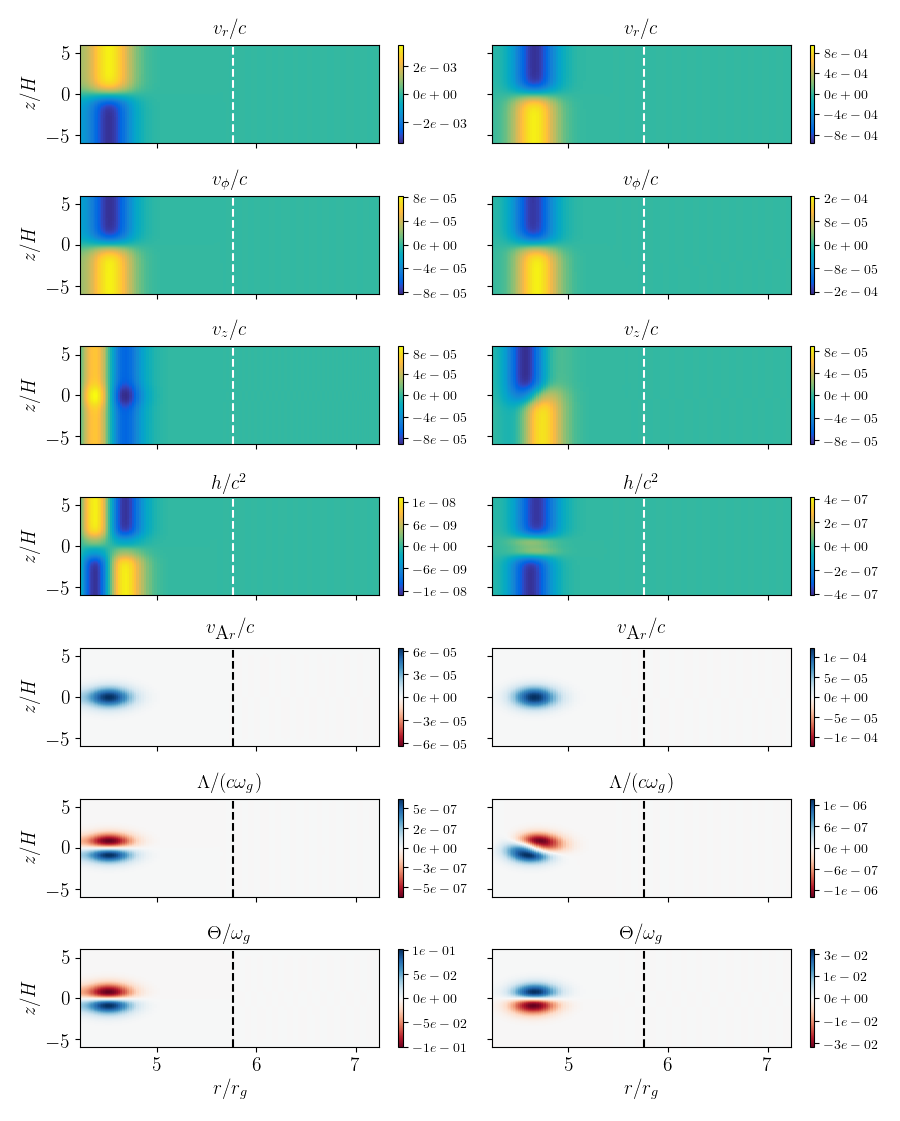}
        \caption{Heatmaps showing the $r,z/H$-dependence for the real parts of the MHD variables of global r-modes calculated with $\beta_z\approx 100$, and both $B_{\phi}=0$ (left-hand column), and $\beta_{\phi}=12.5$ (right-hand column) ($c_s=0.001c,a=0.5,H=H(r_{\textrm{ISCO}})$). The boundary condition $\del_r\delta B_r=0$ is imposed at $r_{\textrm{ISCO}}$, while a wave propagation boundary condition is imposed at $r_{out}$ to account for weak coupling with a 2D inertial-acoustic component. The white and black dashed lines mark the radius of maximum epicyclic frequency.}
        \label{fig:fglob_kzcst}
\end{figure*}

Scaling velocities by $c,$ lengths by $r_g,$ frequencies and $\Theta$
by $\omega_g=c^3/(GM),$ and $\Lambda$ by $c\omega_g$, we solve the
vertically stratified Equations (\ref{eq:strat1})-(\ref{eq:strat7})
using the same hybrid pseudospectral-Galerkin method as in DLO (see
Appendix for details). The vertical and radial structures are more
strongly coupled with the inclusion of a toroidal magnetic field
component. However for moderate azimuthal magnetic field strengths we
find converged solutions using this numerical method. Each calculation
produces a spectrum of r-modes with spatial structure discrete in both
$r$ and $z$, which we characterize with the vertical quantum numbers
$l$ and $n$ (resp.) that best describe the radial velocity
perturbation. Example heatmaps illustrating the $r-z$ dependence of
the fundamental $l=0,n=1$ mode are shown in
Fig. \ref{fig:fglob_kzcst}, calculated with 
$\beta_z=100$, and both $B_{\phi}=0$ (left), and $\beta_{\phi}=12.5$ (right).

Fig. \ref{fig:fglob_kzcst} illustrates the azimuthal magnetic field's
modification of the trapped inertial modes. To begin, there is a mild
shift of the mode localisation outward in the disc, in accord with our
previous calculations. This shift is less dramatic than that shown in Fig. \ref{fig:cyl_ur_hh} because of the weaker $\beta_\phi=12.5$. Most noticeable, however, is the transformation of the modes' magneto-acoustic properties. While in the case of a purely constant vertical field the global r-modes have even or odd symmetry about the mid-plane, with an azimuthal magnetic field of strength $\beta_{\phi}=12.5$ the vertical velocity, enthalpy and Lorentz force perturbations become asymmetric in $z$. This breaking of symmetry for the oscillations is not surprising, since the background magnetic field considered is helical, and asymmetric with respect to the mid-plane. 

The numerical method used in DLO loses utility for strong toroidal fields with
$\beta_{\phi}\lesssim \mathcal{O}(10)$, the regime in which we found the effects of toroidal
magnetic fields on epicyclic-Alfv\'enic r-modes to be most pronounced in
Sections \ref{sec:numdisp} and \ref{sec:cglobal}. This is likely due to the alteration of the magneto-acoustic nature of the r-modes by the toroidal field. The Alfv\'enic restoring force provided by a purely vertical magnetic field does little to change the nearly incompressible nature of hydrodynamic r-modes, and so we found in DLO that in the presence of such a field MHD r-modes' vertical structure is well-described by basis functions derived as the eigenfunctions of anelastic MRI channel modes by \citet{lat10}. As the inertial waves become more compressible, it is natural that their vertical structure can no longer be as closely associated with that of essentially incompressible MRI modes.

Importantly, however, the lengthscale of variation (with respect to the scale height) does not appear to change significantly with increasing $B_{\phi}$, and the frequencies are still close to the predictions made with cylindrical calculations and the $k_z$
prescription $k_z=K_1/H$. This suggests that the calculations made in
Sections \ref{sec:local} and \ref{sec:cglobal} should provide a
reliable window into the behavior of MHD r-modes in the presence of strong toroidal magnetic fields.

\section{Discussions}\label{sec:disc}
In this section we provide theoretical and observational context for our findings. All of the theoretical models offered to explain HFQPOs in black hole binaries thus far face significant difficulties. In describing only particles or single fluid elements, the original 3:2 resonance models of \citet{Abr01} and \citet{Kluz01} result in a very narrow radial extent for the mechanism, which would be unlikely to produce significant changes in emissivity \citep{Rez03}. In applying particle dynamics to a description of a relativistic plasma, the relativistic precession model faces a similar issue, although the theory has been extended to describe the precession of a rigid disc \citep{Mot17}.

With regard to the related models involving the oscillations of accretion tori,
it should first be stated that the geometry of the disc in the SPL state is uncertain. A torus
might be thought of as a proxy for a hot, thick flow interior to a
standard thin disc that is truncated far from the ISCO \citep{Frag16}. But measurements from reflection spectra suggest that at least for GRS 1915+105, the disc truncation radius remains stationary at the ISCO during transitions in and out of high flux emission states \citep[e.g.,][]{Zog16}. Further, oscillations in a torus are likely to be damped by both the Papaloizou-Pringle and magnetorotational instabilities, and to have frequencies sensitive to the torus's non-Keplerian angular momentum distribution \citep{Frag05,Frag16}. Torus oscillations finally lack a convincing mechanism for their excitation to large amplitudes.

Although we have shown in this work that the effects of magnetic
fields on trapped inertial waves may not be as detrimental as
originally thought, diskoseismic oscillations also face challenges in
their explanation of black hole HFQPOs. For one thing, the theory
requires a geometrically thin disc extending nearly to the ISCO. 
Additionally, even if axisymmetric r-modes are robustly excited by disc deformations, the linear theory does not in isolation explain the appearance of HFQPOs in pairs with frequencies in near-integer ratios. However, as mentioned by \citet{ReM09}, there are a number of physical processes not related to resonant phenomenon that might give rise to integer ratios. 

Indeed, \citet{Rem06} noted that in three of the sources exhibiting 3:2 ratios the upper and lower HFQPOs occur in states of weaker and stronger power-law flux (resp.). As suggested by \citet{Fer10}, this might indicate that separate but correlated mechanisms, one of which could be r-mode excitation, are responsible for each oscillation. Alternatively, simulations or a dynamical systems approach might reveal a non-linear interaction between different diskoseismic modes that could give rise to oscillations with frequencies in near-integer ratios, as was preliminarily investigated with a toy model by \citet{ort+14}.

From an observational standpoint, the alteration of r-mode frequencies by poloidal magnetic fields presents an additional challenge in probing the black hole's spin angular momentum. However, the results of this work suggest that the enhancements shown in DLO can be considered upper bounds, as the strong toroidal magnetic field component generated by any net vertical field will drive the frequencies back toward hydrodynamic values. Further, due to an association with magnetic pressure, the effects of a strong toroidal field may translate more readily from our idealized model to a realistic accretion disc than the effects of a vertical field (which are associated with magnetic tension): the MRI, in the presence of a strong ordered vertical flux, generates even stronger fluctuating fields (both toroidal and vertical). Both the toroidal and vertical components of the spatially and temporally varying field will provide an effective magnetic pressure. However, the fluctuating vertical magnetic field is unlikely to provide the same magnetic tension as an ordered vertical field of the equivalent strength, as its spatial and temporal coherence will be limited.

It is worth noting that the detrimental effects of magnetic fields on r-modes may actually prove useful in bridging the gap between theory and observations. A problem that perhaps all models must contend with is the fragility of the HFQPO mechanism. Considering that the features appear only in a handful of sources, and only during specific emission states, most models for HFQPOs are too robust. Trapped inertial waves, for example, might also be expected to provide a signal in the classical high/soft state, when the thin disc unambiguously connects to the ISCO. Conversely, the oscillations of a hot torus might be expected to drive HFQPOs in the low/hard state, when the thin disc is thought to truncate well before the ISCO. In the former case, \citet{FoG09} provided a differentiating factor with their investigation of warp and eccentricity propagation. They found that such deformations are much more likely to overcome damping and travel to the inner disc regions (a requirement for r-mode excitation) when accretion rates are high, close to Eddington. However, magnetic fields might still provide some explanation for the paucity of occurrences across multiple sources.

Finally, throughout we have discussed (vertical) magnetic fields only negatively, solely as an impediment to the theory. But an ordered poloidal magnetic field might be viewed more positively if recognised as a means by which the power in coherent disc oscillations could be transferred to plasma in the corona. It is the corona, after all, that is associated with the high energy `tail' in which the HFQPOs are observed \citep{Done07}. Trapped inertial waves will drive oscillations in any vertical field threading the disc, which on penetrating the coronal plasma might in turn drive activity in its tangled fields. In particular, reconnection driven by the interaction of r-modes with the corona through a vertical magnetic field could lead to time periodic dissipation and hence emission. This is an idea worth exploring, and permits the extension of the theory from its current, purely dynamical form.

If trapped inertial wave excitation proves a robust explanation for HFQPOs, the features might offer a window into the presence (or lack thereof) of strong magnetic fields in systems for which the spin is already constrained. However, the controlled analytical and numerical experiments presented in this work constitute only one piece of the puzzle. Development of the theory to include non-linearity and more complete treatments of vertical structure and both radiative and thermal physics are required before observations can be confronted with certainty.

\section{Conclusions}\label{sec:conc}
We have explored the effects of large scale magnetic fields with both
toroidal and poloidal components on trapped inertial waves (r-modes)
in MHD models of relativistic accretion discs, through both local
analyses and global normal mode calculations. Previous studies
\citep{FL09,DLO} suggested that purely azimuthal magnetic fields affect r-mode trapping minimally. However, we find that when considered in conjunction with, rather than in isolation from, a poloidal field component, toroidal magnetic fields have a greater impact. Far from remaining passive, a background azimuthal magnetic field reduces the effects of a vertical one on trapped inertial waves, moving r-mode trapping regions and frequencies back toward their hydrodynamic values. This finding is not in opposition to \citet{FL09} or \citet{DLO}; the purely azimuthal and purely vertical magnetic fields considered in those works can be seen as special cases of the field configurations investigated here.

Quantitatively, for any toroidal field strength the radial geometry of inertial wave trapping is only marginally affected by weak to moderately strong vertical magnetic fields with mid-plane plasma betas $\beta_z\gtrsim 500$. The detrimental effects of stronger vertical fields are greatly reduced by azimuthal fields with $B_{\phi}\gg B_z$, and the isolation of the trapping cavity from the ISCO can even be restored for $\beta_z\lesssim200$ by  toroidal magnetic field components contributing $\beta_{\phi}\lesssim 10$. Global and local MHD simulations suggest these conditions are not unreasonable near the disc mid-plane \citep[e.g.,][]{salv16a,Zhu18}, although the vertical profile for the azimuthal magnetic field will certainly be more complicated than that considered here. 

Further, we have shown that the effect of a strong azimuthal component of a helical magnetic field is to modify r-modes' magneto-acoustic nature, while a purely vertical field modifies r-modes primarily through a restoring force due to magnetic tension. 
This suggests that the restorative effects of a strong toroidal component will translate easily from our simplified model assuming smooth, large-scale magnetic fields to a more realistic disc with magnetic fields that are disordered. 

The model presented in this paper is necessarily idealised, ignoring the effects of radial inflow and radiation pressure, both of which ought to be relevant in the emission states in which HFQPOs are observed. The vertical structure of both the disc and the background toroidal field itself should also be considered more carefully, as the latter will significantly impact the former for plasma betas approaching or surpassing thermal strengths \citep{terp96}. However, the qualitative result that background toroidal magnetic fields reduce the detrimental effects of vertical magnetic fields on trapped inertial waves should translate to more complicated treatments of vertical structure. The survival of trapped inertial waves in a realistic accretion flow is most likely to be determined by a competition between excitation by global disc deformations and damping by radial inflow and turbulent fluctuations. More sophisticated analytical and numerical models will be required to determine if r-modes can provide a robust explanation for HFQPOs. 

\section*{Acknowledgements}
The authors thank the anonymous reviewer for helpful comments and suggestions that significantly improved the paper. J. Dewberry thanks the Cambridge Commonwealth European and International Trust and the Vassar College De Golier Trust for funding this work.



\bibliographystyle{mnras}
\bibliography{strongtor} 

\begin{thebibliography}{}
\makeatletter
\relax
\def\mn@urlcharsother{\let\do\@makeother \do\$\do\&\do\#\do\^\do\_\do\%\do\~}
\def\mn@doi{\begingroup\mn@urlcharsother \@ifnextchar [ {\mn@doi@}
  {\mn@doi@[]}}
\def\mn@doi@[#1]#2{\def\@tempa{#1}\ifx\@tempa\@empty \href
  {http://dx.doi.org/#2} {doi:#2}\else \href {http://dx.doi.org/#2} {#1}\fi
  \endgroup}
\def\mn@eprint#1#2{\mn@eprint@#1:#2::\@nil}
\def\mn@eprint@arXiv#1{\href {http://arxiv.org/abs/#1} {{\tt arXiv:#1}}}
\def\mn@eprint@dblp#1{\href {http://dblp.uni-trier.de/rec/bibtex/#1.xml}
  {dblp:#1}}
\def\mn@eprint@#1:#2:#3:#4\@nil{\def\@tempa {#1}\def\@tempb {#2}\def\@tempc
  {#3}\ifx \@tempc \@empty \let \@tempc \@tempb \let \@tempb \@tempa \fi \ifx
  \@tempb \@empty \def\@tempb {arXiv}\fi \@ifundefined
  {mn@eprint@\@tempb}{\@tempb:\@tempc}{\expandafter \expandafter \csname
  mn@eprint@\@tempb\endcsname \expandafter{\@tempc}}}

\bibitem[\protect\citeauthoryear{Abramowicz \& Kl\'uzniak}{Abramowicz \&
  Kl\'uzniak}{2001}]{Abr01}
Abramowicz M.~A.,  Kl\'uzniak W.,  2001, A{\&}A, 374, 19

\bibitem[\protect\citeauthoryear{Bai \& Stone}{Bai \& Stone}{2013}]{Bai13}
Bai X.-N.,  Stone J.~M.,  2013, ApJ, 767, 30

\bibitem[\protect\citeauthoryear{Belloni, Sanna  \& Mendez}{Belloni
  et~al.}{2012}]{Bel12}
Belloni T.~M.,  Sanna A.,   Mendez M.,  2012, MNRAS, 426, 1701

\bibitem[\protect\citeauthoryear{Blaes, Arras  \& Fragile}{Blaes
  et~al.}{2006}]{Blaes06}
Blaes O.~M.,  Arras P.,   Fragile P.~C.,  2006, MNRAS, 369, 1235

\bibitem[\protect\citeauthoryear{Boyd}{Boyd}{2001}]{boyd}
Boyd J.~P.,  2001, Chebyshev and Fourier spectral methods, 2 edn.
Dover Publications, Mineola NY

\bibitem[\protect\citeauthoryear{Cabanac, Henri, Petrucci, Malzac, Ferreira  \&
  Belloni}{Cabanac et~al.}{2010}]{Cab}
Cabanac C.,  Henri G.,  Petrucci P.-O.,  Malzac J.,  Ferreira J.,   Belloni
  T.~M.,  2010, MNRAS, 404, 738

\bibitem[\protect\citeauthoryear{Dewberry, Latter  \& Ogilvie}{Dewberry
  et~al.}{2018}]{DLO}
Dewberry J.~W.,  Latter H.~N.,   Ogilvie G.~I.,  2018, MNRAS, 476, 4085

\bibitem[\protect\citeauthoryear{Dexter \& Blaes}{Dexter \&
  Blaes}{2014}]{DexBl}
Dexter J.,  Blaes O.,  2014, MNRAS, 438, 3352

\bibitem[\protect\citeauthoryear{Done, Gierl\'inski  \& Kubota}{Done
  et~al.}{2007}]{Done07}
Done C.,  Gierl\'inski M.,   Kubota A.,  2007, Astron. Astrophys. Rev., 15, 1

\bibitem[\protect\citeauthoryear{Ferreira}{Ferreira}{2010}]{Fer10}
Ferreira B.,  2010, PhD thesis, University of Cambridge

\bibitem[\protect\citeauthoryear{Ferreira \& Ogilvie}{Ferreira \&
  Ogilvie}{2008}]{FoG08}
Ferreira B.~T.,  Ogilvie G.~I.,  2008, MNRAS, 386, 2297

\bibitem[\protect\citeauthoryear{Ferreira \& Ogilvie}{Ferreira \&
  Ogilvie}{2009}]{FoG09}
Ferreira B.~T.,  Ogilvie G.~I.,  2009, MNRAS, 392, 428

\bibitem[\protect\citeauthoryear{Fragile}{Fragile}{2005}]{Frag05}
Fragile P.~C.,  2005, Proc. of the 22nd Texas Symposium on Relativistic
  Astrophysics, pp 270--275

\bibitem[\protect\citeauthoryear{Fragile, Straub  \& Blaes}{Fragile
  et~al.}{2016}]{Frag16}
Fragile C.,  Straub O.,   Blaes O.,  2016, MNRAS, 461, 1356

\bibitem[\protect\citeauthoryear{Fu \& Lai}{Fu \& Lai}{2009}]{FL09}
Fu W.,  Lai D.,  2009, ApJ, 690, 1386

\bibitem[\protect\citeauthoryear{Hawley, Gammie  \& Balbus}{Hawley
  et~al.}{1995}]{Haw95}
Hawley J.~F.,  Gammie C.~F.,   Balbus S.~A.,  1995, ApJ, 440, 742

\bibitem[\protect\citeauthoryear{Hor{\'{a}}k}{Hor{\'{a}}k}{2008}]{Horak08}
Hor{\'{a}}k J.,  2008, A{\&}A, 486, 1

\bibitem[\protect\citeauthoryear{Kato}{Kato}{2001}]{kat01}
Kato S.,  2001, PASJ, 53, 1

\bibitem[\protect\citeauthoryear{Kato}{Kato}{2004}]{Kat04}
Kato S.,  2004, PASJ, 56, 905

\bibitem[\protect\citeauthoryear{Kato}{Kato}{2008}]{Kat08}
Kato S.,  2008, PASJ, 60, 111

\bibitem[\protect\citeauthoryear{Kersale, Hughes, Ogilvie, Tobias  \&
  Weiss}{Kersale et~al.}{2004}]{Ker04}
Kersale E.,  Hughes D.~W.,  Ogilvie G.~I.,  Tobias S.~M.,   Weiss N.~O.,  2004,
  ApJ, 602, 892

\bibitem[\protect\citeauthoryear{Kl\'uzniak \& Abramowicz}{Kl\'uzniak \&
  Abramowicz}{2001}]{Kluz01}
Kl\'uzniak W.,  Abramowicz M.~A.,  2001, Acta Physica Polonica B, 32, 3605

\bibitem[\protect\citeauthoryear{Latter, Fromang  \& Gressel}{Latter
  et~al.}{2010}]{lat10}
Latter H.~N.,  Fromang S.,   Gressel O.,  2010, MNRAS, 406, 848

\bibitem[\protect\citeauthoryear{Li, Goodman  \& Narayan}{Li
  et~al.}{2003}]{Li02}
Li L.-X.,  Goodman J.,   Narayan R.,  2003, ApJ, 593, 980

\bibitem[\protect\citeauthoryear{Miller et~al.,}{Miller et~al.}{2016}]{Mill16}
Miller J.~M.,  et~al., 2016, ApJ Letters, 821, 9

\bibitem[\protect\citeauthoryear{Morgan, Remillard  \& Greiner}{Morgan
  et~al.}{1997}]{Morg97}
Morgan E.~H.,  Remillard R.~A.,   Greiner J.,  1997, ApJ, 482, 993

\bibitem[\protect\citeauthoryear{Motta}{Motta}{2016}]{Mot16}
Motta S.~E.,  2016, Astron. Nachr., 337, 398

\bibitem[\protect\citeauthoryear{Motta, Belloni, Stella, Mu{\~{n}}oz-Darias  \&
  Fender}{Motta et~al.}{2014a}]{Mot14}
Motta S.~E.,  Belloni T.~M.,  Stella L.,  Mu{\~{n}}oz-Darias T.,   Fender R.,
  2014a, MNRAS, 437, 2554

\bibitem[\protect\citeauthoryear{Motta, Mu{\~{n}}oz-Darias, Sanna, Fender,
  Belloni  \& Stella}{Motta et~al.}{2014b}]{Mot14b}
Motta S.~E.,  Mu{\~{n}}oz-Darias T.,  Sanna A.,  Fender R.,  Belloni T.,
  Stella L.,  2014b, MNRAS, 439, 65

\bibitem[\protect\citeauthoryear{Motta, Franchini, Lodato  \&
  Mastroserio}{Motta et~al.}{2018}]{Mot17}
Motta S.~E.,  Franchini A.,  Lodato G.,   Mastroserio G.,  2018, MNRAS, 473,
  431

\bibitem[\protect\citeauthoryear{Nayakshin, Rappaport  \& Melia}{Nayakshin
  et~al.}{2000}]{Nay0}
Nayakshin S.,  Rappaport S.,   Melia F.,  2000, ApJ, 535, 798

\bibitem[\protect\citeauthoryear{Ogilvie \& Pringle}{Ogilvie \&
  Pringle}{1996}]{Og96}
Ogilvie G.~I.,  Pringle J.~E.,  1996, MNRAS, 279, 152

\bibitem[\protect\citeauthoryear{Okazaki, Kato  \& Fukue}{Okazaki
  et~al.}{1987}]{Oka87}
Okazaki A.,  Kato S.,   Fukue J.,  1987, PASJ, 39, 457

\bibitem[\protect\citeauthoryear{Ortega-Rodr\'iguez, Sol\'is-S\'anchez,
  L\'opez-Barquero, Matamoros-Alvarado  \& Venegas-Li}{Ortega-Rodr\'iguez
  et~al.}{2014}]{ort+14}
Ortega-Rodr\'iguez M.,  Sol\'is-S\'anchez H.,  L\'opez-Barquero V.,
  Matamoros-Alvarado B.,   Venegas-Li A.,  2014, MNRAS, 440, 3011

\bibitem[\protect\citeauthoryear{Papaloizou \& Pringle}{Papaloizou \&
  Pringle}{1984}]{Papri}
Papaloizou J. C.~B.,  Pringle J.~E.,  1984, MNRAS, 208, 721

\bibitem[\protect\citeauthoryear{Remillard \& McClintock}{Remillard \&
  McClintock}{2006}]{Rem06}
Remillard R.~A.,  McClintock J.~E.,  2006, Ann. Rev. Astron. Astrophys., 44, 49

\bibitem[\protect\citeauthoryear{Reynolds \& Miller}{Reynolds \&
  Miller}{2009}]{ReM09}
Reynolds C.~S.,  Miller M.~C.,  2009, ApJ, 692, 869

\bibitem[\protect\citeauthoryear{Rezzolla, Yoshida, Maccarone  \&
  Zanotti}{Rezzolla et~al.}{2003}]{Rez03}
Rezzolla L.,  Yoshida S.,  Maccarone T.,   Zanotti O.,  2003, MNRAS, 344, 37

\bibitem[\protect\citeauthoryear{Salvesen, Simon, Armitage  \&
  Begelman}{Salvesen et~al.}{2016}]{salv16a}
Salvesen G.,  Simon J.~B.,  Armitage P.~J.,   Begelman M.~C.,  2016, MNRAS,
  457, 857

\bibitem[\protect\citeauthoryear{Sano \& Miyama}{Sano \& Miyama}{1999}]{sam99}
Sano T.,  Miyama S.~M.,  1999, ApJ, 515, 776

\bibitem[\protect\citeauthoryear{Stella \& Vietri}{Stella \&
  Vietri}{1998}]{Stel98}
Stella L.,  Vietri M.,  1998, ApJ, 492, 59

\bibitem[\protect\citeauthoryear{Stella, Vietri  \& Morsink}{Stella
  et~al.}{1999}]{Stel99}
Stella L.,  Vietri M.,   Morsink S.~M.,  1999, ApJ, 524, 63

\bibitem[\protect\citeauthoryear{Stone, Hawley, Gammie  \& Balbus}{Stone
  et~al.}{1996}]{Sto96}
Stone J.~M.,  Hawley J.~F.,  Gammie C.~F.,   Balbus S.~A.,  1996, ApJ, 463, 656

\bibitem[\protect\citeauthoryear{Terquem \& Papaloizou}{Terquem \&
  Papaloizou}{1996}]{terp96}
Terquem C.,  Papaloizou J. C.~B.,  1996, MNRAS, 279, 767

\bibitem[\protect\citeauthoryear{Wagoner}{Wagoner}{2012}]{Wag12}
Wagoner R.,  2012, ApJ Letters, 752, 18

\bibitem[\protect\citeauthoryear{Zhu \& Stone}{Zhu \& Stone}{2018}]{Zhu18}
Zhu Z.,  Stone J.~M.,  2018, ApJ, 857, 34

\bibitem[\protect\citeauthoryear{Zoghbi et~al.,}{Zoghbi et~al.}{2016}]{Zog16}
Zoghbi A.,  et~al., 2016, ApJ, 833, 165

\makeatother
\end{thebibliography}


\appendix
\twocolumn[
\section{Numerical method for density stratification}
To solve Equations (\ref{eq:lowE1})-(\ref{eq:lowE7}) for fully global r-mode solutions, we first trade variables, as described in Section 4.2. We additionally make a change of coordinate, exchanging $z$ for $\eta(r,z)=z/H(r)$. Retaining radial variation in the scale height, the full set of equations is then 
\begin{align}
\label{eq:strat1}
    -\ti\omega v_r\ 
    &=2\Omega v_{\phi}
    -c_s^2
    \left(
        \dfrac{\del }{\del r} 
        -\dfrac{\td\ln H}{\td r}\dfrac{\del }{\del \ln \eta}
    \right)\Gamma
    +\dfrac{1}{g}\Lambda,
\\
    -\ti\omega v_{\phi}\ 
    &=-\dfrac{\kappa^2}{2\Omega}v_r 
    +\dfrac{V_{\tA z}}{g}\Theta,
\\
    -\ti\omega v_z\ 
    &=-\dfrac{c_s^2}{H}\dfrac{\del \Gamma}{\del \eta }
    -\dfrac{V_{\tA\phi}}{g}\Theta,
\\
    -\ti\omega \Gamma\ \ 
    &=-\dfrac{1}{r}\dfrac{\del (rv_r)}{\del r}
    +\dfrac{\td\ln H}{\td r}\dfrac{1}{g}\dfrac{\del (g v_r)}{\del\ln\eta}
    -\dfrac{1}{gH}\dfrac{\del (gv_z)}{\del \eta},
\\
    -\ti\omega v_{\tA r}
    &=\dfrac{V_{\tA z}}{H}\dfrac{\del v_r}{\del \eta},
\\\label{eq:strat6}
    -\ti \omega \Lambda\ 
    &=\left\{
        \mathcal{L}_{\tA}
        +\dfrac{1}{H}\mathcal{L}_H\dfrac{\del }{\del \ln \eta}
        +\dfrac{1}{H^2}
        \left(
            V_{\tA z}^2
            +V_{\tA}^2\left(\del_rH\right)^2\eta^2
        \right)
        \dfrac{\del^2}{\del \eta^2}
    \right\}v_r
    -V_{\tA\phi}V_{\tA z}
    \left\{
        \dfrac{1}{r}\dfrac{\del }{\del r}
        \left(\dfrac{r}{H}\dfrac{\del v_{\phi}}{\del \eta}\right)
        -\dfrac{1}{H}\dfrac{\td\ln H}{\td r}
        \eta\dfrac{\del^2 v_{\phi}}{\del \eta^2}
    \right\}
\\&\hspace{4cm}\notag 
    +V_{\tA\phi}^2
    \left\{
        \dfrac{\del }{\del r}\left(\dfrac{1}{H}\dfrac{\del v_z}{\del \eta}\right)
        -\dfrac{1}{H}\dfrac{\td\ln H}{\td r}\eta\dfrac{\del^2 v_z}{\del \eta^2}
    \right\}
    -V_{\tA\phi}
    \left(
        \mathcal{L}_{\Omega} 
        -\dfrac{\td\Omega}{\td \ln r}\dfrac{\td\ln H}{\td r}
        \dfrac{\del }{\del\ln \eta}
    \right)v_{\tA r},
\\\label{eq:strat7}
-\ti \omega \Theta\ 
    &=-V_{\tA\phi}\dfrac{r}{H}
    \dfrac{\del^2 }{\del \eta \del r}\left(\dfrac{v_r}{r}\right)
    +\dfrac{V_{\tA z}}{H^2}\dfrac{\del^2 v_{\phi}}{\del \eta^2}
    -\dfrac{V_{\tA\phi}}{H^2}\dfrac{\del^2 v_z}{\del \eta^2}
    +\dfrac{1}{H}\dfrac{d\Omega}{d \ln r}\dfrac{\del v_{\tA r}}{\del \eta},
\end{align}
where we have defined the purely radial differential operators
\begin{align}
    \mathcal{L}_{\tA}&=
    (V_{\tA\phi}^2+V_{\tA z}^2)\dfrac{\del^2}{\del r^2}
    +\dfrac{1}{r}(V_{\tA z}^2-V_{\tA\phi}^2)
    \left(
        \dfrac{\del }{\del r}
        -\dfrac{1}{r}
    \right),
\\
    \mathcal{L}_H
    &=\dfrac{1}{r}\left(V_{\tA\phi}^2-V_{\tA z}^2\right)\dfrac{\td H}{\td r}
    -V_{\tA}^2
    \left\{
        2\dfrac{\td  H}{\td r}\dfrac{\del}{\del r}
        +H^2\dfrac{\td }{\td r}
        \left(
            \dfrac{1}{H^2}\dfrac{\td H}{\td r}
        \right)
    \right\},
\\
    \mathcal{L}_{\Omega}
    &=\dfrac{\td\Omega}{\td \ln r}\dfrac{\del }{\del r}
    +\dfrac{1}{r}\dfrac{\td }{\td r}\left(r\dfrac{\td\Omega}{\td \ln r}\right).
\end{align}
We expand the perturbations as 
\begin{align}\label{eq:exp1}
	v_r(r,\eta)&=\sum_{m=0}^{\infty}u_m(r)F_m(\eta),\qquad
	v_{\phi}(r,\eta)=\sum_{m=0}^{\infty}v_m(r)F_m(\eta),\qquad
	w_r(r,\eta)=\sum_{m=1}^{\infty}w_m(r)F_{m-1}(\eta),\\
	v_{\tA r}(r,\eta)&=\sum_{m=0}^{\infty}v_{\tA}^m(r)G_m(\eta),\qquad
	\Lambda(r,\eta)=\sum_{m=0}^{\infty}\Lambda_m(r)G_m(\eta),\qquad
	\Theta(r,\eta)=\sum_{m=0}^{\infty}\Theta_m(r)g(\eta)F_m(\eta),  \label{eq:exp7}
\end{align}
where $\{F_n(\eta)\}$ and $\{G_n(\eta)\}$ are orthogonal sets of dimensionless basis functions characterized by a vertical quantum number $n$ and corresponding eigenvalue $K_n$. They describe the vertical structure of the horizontal velocity and magnetic field components (resp.) of magnetorotational channel modes in the anelastic approximation \citep{lat10}, and can be normalised such that 
\begin{align}
\label{eq:orth1}
    \int_{-\infty}^{\infty}gF_nF_m\td\eta=\delta_{nm},\qquad
    \int_{-\infty}^{\infty}G_nG_m\td\eta=\delta_{nm}.
\end{align}
Since the $F_n$ go to a constant as $\eta\rightarrow\infty$ and $G_n$ to zero, the expansions (\ref{eq:exp1})-(\ref{eq:exp7}) implicitly place boundary conditions on the perturbation variables' behaviors far above and below the disc; namely, we assume that in the rarefied, magnetically dominated atmosphere the velocity components and $\Gamma=\delta\rho/\rho$ are stabilized and forced to go to constants. For the true Alfv\'en velocity perturbation $v_{\tA r}/\sqrt{g}$ to remain bounded $v_{\tA r}$ must go to zero. With regard to the Lorentz force perturbations, we adopt a hot halo model \citep{sam99} and assume that far from the mid-plane the magnetic field achieves a force-free configuration, such that $\Lambda$ and $\Theta\rightarrow 0$.
]

\twocolumn[
Substituting Expansions (\ref{eq:exp1})-(\ref{eq:exp7}) into Equations (\ref{eq:strat1})-(\ref{eq:strat7}), and using the orthogonality relations (\ref{eq:orth1}) to project onto an arbitrary vertical order $n$ leaves the coupled equations
\begin{align}
\label{eq:strat_coup1}
    -\ti\omega u_n
    &=2\Omega v_n
    -c_s^2\dfrac{\td \Gamma_n}{\td r} 
    +c_s^2\del_rH\sum_{m=0}^{\infty}
    \hat{K}_m\nu_{nm}\Gamma_m
    +\sum_{m=0}^{\infty}\mu_{nm}\Lambda_m,
\\
    -\ti\omega v_n\ 
    &=-\dfrac{\kappa^2}{2\Omega}u_n
    +V_{\tA z}\Theta_n ,  
\\
    -\ti\omega w_n
    &=-c_s^2\sum_{m=0}^{\infty}\hat{K}_m\epsilon_{n-1,m} \Gamma_m
    -V_{A\phi}\Theta_{n-1},
\\
    -\ti\omega \Gamma_n
    &=-\dfrac{1}{r}\dfrac{\td(ru_n)}{\td r}
    -\del_r\ln H\sum_{m=0}^{\infty}\lambda_{nm}u_m
    +\hat{K}_n\sum_{m=1}^{\infty}\epsilon_{m-1,n}w_m,
\\
    -\ti\omega v_{\tA}^n
    &=\hat{K}_nV_{\tA z}u_n,
\\\notag \notag 
    -\ti\omega \Lambda_n
    &=\sum_{m=0}^{\infty}
    \left\{
        \mu_{mn}\mathcal{L}_A
        +\hat{K}_m\gamma_{nm}\mathcal{L}_H
        -\hat{K}_m^2
        \left(
            V_{\tA z}^2\epsilon_{mn}
            +V_{\tA}^2\left(\del_r H\right)^2\alpha_{mn}
        \right)
    \right\}u_m
    -V_{\tA\phi}V_{Az}
    \left(
        \dfrac{1}{r}\dfrac{\td }{\td r}
        \left(\hat{K}_nrv_n\right)
        +\del_r H\sum_{m=0}^{\infty}\hat{K}_m^2\nu_{mn}v_m
    \right)
\\&
    +V_{\tA\phi}^2
    \left(
        \dfrac{\td }{\td r}\left(\hat{K}_nw_{n+1}\right)
        +\del_rH\sum_{m=1}^{\infty}\hat{K}_{m-1}^2\nu_{m-1,n}w_m
    \right)
    -V_{\tA\phi}
    \left(
        \mathcal{L}_{\Omega}v_{\tA}^n
        +\dfrac{\td\Omega}{\td \ln r}\del_r H\sum_{m=0}^{\infty}\hat{K}_m\nu_{mn}v_{\tA}^m 
    \right),
\\
\label{eq:strat_coup7}
    -\ti\omega \Theta_n
    &=-V_{\tA\phi}\sum_{m=0}^{\infty}\hat{K}_m\mu_{nm}
    \dfrac{\td }{\td \ln r}\left(\dfrac{u_m}{r}\right)
    -V_{\tA z}\hat{K}_n^2v_n
    +V_{\tA\phi}\hat{K}_nw_{n+1}
    -\hat{K}_n\dfrac{d\Omega}{d \ln r}v_{\tA}^n,
\end{align}
where the sums are over finite coupling integrals defined by
\begin{align}
    \mu_{nm}&=\int_{-\infty}^{\infty}F_nG_m\td\eta,\ \ \qquad
    \epsilon_{nm}=\int_{-\infty}^{\infty}gF_nG_m\td\eta,\ \ \qquad
    \lambda_{nm}=\int_{-\infty}^{\infty}g F_m\left(F_n+\eta K_nG_n\right)\td\eta,\\
    \gamma_{nm}&=\int_{-\infty}^{\infty}\eta G_nG_m\td\eta,\qquad
    \nu_{nm}=\int_{-\infty}^{\infty}\eta gF_nG_m\td\eta,\qquad
    \alpha_{nm}=\int_{-\infty}^{\infty}\eta^2gF_nG_m\td\eta.
\end{align}

The equations are most heavily coupled by the sums in Equations (\ref{eq:strat_coup1}) and (\ref{eq:strat_coup7}) that involve the integrals $\mu_{nm}$, which increase in importance with increasing $B_{\phi}.$ As a consequence of the symmetry breaking by the mixed magnetic field components, the vertically structured r-modes are coupled to 2D, $n=0$ components. This coupling, which cannot be described by a cylindrical model in which the vertical wavenumber is either zero or non-zero, allows for some wave leakage out of the trapping region (this leakage is treated with a wave propagation boundary condition, although the decay rates it introduces are minimal). 

\vspace{1em}
Despite this stronger coupling, solving Equations (\ref{eq:strat_coup1})-(\ref{eq:strat_coup7}) as a generalized eigenvalue problem for the radially varying coefficients and reconstructing the full solutions from Equations (\ref{eq:exp1})-(\ref{eq:exp7}) (see DLO for more details) produces frequencies and eigenmodes that converge with truncation of the summation terms at increasingly large $m=M$ (see Tables \ref{tab:cstkz_converge} and \ref{tab:varkz_converge}). However, for strong $B_{\phi}$ with $\beta_{\phi}\lesssim10$ this convergence is too slow for this pseudospectral-Galerkin method to be useful, as the toroidal magnetic field changes the r-modes' compressibility and renders the $\{F_n\}$ and $\{G_n\}$ basis functions innapropriate.
\vspace{2em}
]
\begin{table*}	
\centering
	\begin{tabular}{lcccccccr} 
		\hline
		$V_{\tA\phi}/c_s $ & 0.1 & 0.15 & 0.2 & 0.25 & 0.3 & 0.35 & 0.4 \\
		\hline
		$\sim\beta_{\phi} $ & 200 & 88.9 & 50 & 32 & 22.2 & 16.3 & 12.5 \\
		\hline
		$M=2$ & 0.04032 & 0.04029 & 0.04025 & 0.04020 & 0.04013 & 0.04006 & 0.03998 \\
		$M=3$ & 0.04033 & 0.04029 & 0.04025 & 0.04020 & 0.04014 & 0.04006 & 0.03998 \\
		$M=4$ & 0.04036 & 0.04033 & 0.04029 & 0.04024 & 0.04018 & 0.04011 & 0.04004 \\
		$M=5$ & 0.04036 & 0.04033 & 0.04029 & 0.04024 & 0.04018 & 0.04011 & 0.04004 \\
		$M=6$ & 0.04037 & 0.04034 & 0.04030 & 0.04025 & 0.04020 & 0.04014 & 0.04007 \\
		$M=7$ & 0.04037 & 0.04034 & 0.04030 & 0.04025 & 0.04020 & 0.04014 & 0.04007 \\
		$M=8$ & 0.04037 & 0.04034 & 0.04031 & 0.04026 & 0.04021 & 0.04015 & 0.04010 \\
		$M=9$ & 0.04037 & 0.04034 & 0.04031 & 0.04026 & 0.04021 & 0.04015 & 0.04009 \\
		$M=10$ & 0.04037 & 0.04035 & 0.04031 & 0.04027 & 0.04022 & 0.04017 & 0.04012 \\
		$M=11$ & 0.04037 & 0.04035 & 0.04031 & 0.04027 & 0.04022 & 0.04017 & 0.04012 \\
		$M=12$ & 0.04037 & 0.04035 & 0.04031 & 0.04027 & 0.04023 & 0.04018 & 0.04014 \\
		$M=13$ & 0.04037 & 0.04035 & 0.04031 & 0.04027 & 0.04023 & 0.04018 & 0.04014 \\
		$M=14$ & 0.04037 & 0.04035 & 0.04032 & 0.04028 & 0.04023 & 0.04019 & 0.04019 \\
		\hline
		Cyl ($k_z=K_1/H$) & 0.04040 & 0.04037 & 0.04032 & 0.04025 & 0.04017 & 0.04007 & 0.03996 \\
		\hline
	\end{tabular}
	\caption{Frequencies of fully global, fundamental r-modes, for $V_{\tA z}/c_s=0.06$ ($\beta_z\approx 555$) and varying values of $\beta_{\phi}$, calculated as solutions to a series of coupled eigenvalue problems with a constant scale-height approximation, truncated at different vertical orders $m=M$ ($c_s=0.002c$, $a=0.5$, $N=150$ Gauss-Lobatto grid-points, $r\in[4.2331,10.2331]r_g$). The bottom row gives the frequencies calculated using the cylindrical model with the same parameters and $k_z=K_1/H(r_{\textrm{ISCO}})$. Modes are calculated with the boundary conditions $\del_r\delta B_r=0$ at $r_\text{in}$ and $\del_rv_r=\ti k_rv_r$ at $r_\text{out}$ (radial wavenumber $k_r$ determined for the $2D,$ $n=0$ component which weakly couples r-modes to inertial-acoustic oscillation outside of the trapping region).}
	\label{tab:cstkz_converge}
\end{table*}

\begin{table*}
\centering
	\begin{tabular}{lccccccccr} 
		\hline
		$V_{\tA\phi}/c_s $ & 0.1 & 0.15 & 0.2 & 0.25 & 0.3 & 0.35 & 0.4 \\
		\hline
		$\sim\beta_{\phi} $ & 200 & 88.9 & 50 & 32 & 22.2 & 16.3 & 12.5 \\
		\hline
		$M=2$ & 0.03768 & 0.03765 & 0.03760 & 0.03755 & 0.03749 & 0.03742 & 0.03733 \\
		$M=3$ & 0.03769 & 0.03765 & 0.03761 & 0.03756 & 0.03749 & 0.03742 & 0.03734 \\
		$M=4$ & 0.03771 & 0.03768 & 0.03764 & 0.03759 & 0.03753 & 0.03746 & 0.03739 \\
		$M=5$ & 0.03771 & 0.03768 & 0.03764 & 0.03759 & 0.03753 & 0.03746 & 0.03738 \\
		$M=6$ & 0.03771 & 0.03769 & 0.03765 & 0.03760 & 0.03754 & 0.03748 & 0.03741 \\
		$M=7$ & 0.03771 & 0.03769 & 0.03765 & 0.03760 & 0.03754 & 0.03748 & 0.03741 \\
		$M=8$ & 0.03772 & 0.03769 & 0.03765 & 0.03761 & 0.03755 & 0.03749 & 0.03743 \\
		$M=9$ & 0.03772 & 0.03769 & 0.03765 & 0.03761 & 0.03755 & 0.03749 & 0.03743 \\
		$M=10$ & 0.03772 & 0.03769 & 0.03766 & 0.03761 & 0.03756 & 0.03750 & 0.03744 \\
		$M=11$ & 0.03772 & 0.03769 & 0.03766 & 0.03761 & 0.03756 & 0.03750 & 0.03744 \\
		$M=12$ & 0.03772 & 0.03769 & 0.03766 & 0.03762 & 0.03757 & 0.03751 & 0.03746 \\
		$M=13$ & 0.03772 & 0.03769 & 0.03766 & 0.03762 & 0.03757 & 0.03751 & 0.03746 \\
		$M=14$ & 0.03772 & 0.03770 & 0.03766 & 0.03762 & 0.03757 & 0.03752 & 0.03747 \\
		\hline
		Cyl ($k_z=K_1/H$) & 0.03776 & 0.03772 & 0.03767 & 0.03760 & 0.03752 & 0.03743 & 0.03732 \\
		\hline
	\end{tabular}
	\caption{Frequencies of fully global, fundamental r-modes, for $V_{\tA z}/c_s=0.06$ ($\beta_z\approx 555$) and varying values of $\beta_{\phi}$, calculated as solutions to a series of coupled eigenvalue problems with the extra coupling provided by a radially varying scale-height, truncated at different vertical orders $m=M$ ($c_s=0.002c$, $a=0.5$, $N=150$ Gauss-Lobatto grid-points, $r\in[4.2331,10.2331]r_g$). The bottom row gives the frequencies calculated using the cylindrical model with the same parameters and $k_z=K_1/H(r)$. Modes are calculated with the boundary conditions $\del_r\delta B_r=0$ at $r_\text{in}$ and $\del_rv_r=\ti k_rv_r$ at $r_\text{out}$ (radial wavenumber $k_r$ determined from the full dispersion relation for non-zero $k_z$, which dominates leakage outside of the trapping region when vertical scale height variation is included).}
	\label{tab:varkz_converge}
\end{table*}


\bsp	
\label{lastpage}
\end{document}